\documentclass{aa}  
\usepackage{natbib}
\usepackage{graphicx}
\usepackage{txfonts}
\usepackage{xcolor}
\usepackage{adjustbox}
\usepackage{threeparttable}
\usepackage{amsmath}
\usepackage{amssymb}
\usepackage[version=4]{mhchem}
\usepackage[colorlinks=true,linkcolor=blue,citecolor=blue,filecolor=blue,urlcolor=blue]{hyperref}
\usepackage{comment}
\usepackage[switch]{lineno}

\definecolor{correc}{RGB}{0, 0, 0}

\begin{document} 
   \title{\textbf{Compositional characterisation of asteroid (84) Klio with JWST}}

   \author{
          Tania Le Pivert-Jolivet
          \inst{1,2}
          \and
          Julia de Le\'on
          \inst{1,2}
          \and
          Javier Licandro
          \inst{1,2}
          \and
          Bryan Holler
          \inst{3}
          \and
          Noem\'i  Pinilla-Alonso
          \inst{4,5}
          \and
          Mário De Pr\'a
          \inst{6}
          \and
          Joshua Emery
          \inst{7}
          \and
          Brittany Harvison
          \inst{5}
          \and
          Joseph Masiero
          \inst{8}
          \and
          Lucas McClure
          \inst{7}
          \and
          Driss Takir\inst{9}
          }

   \institute{Instituto de Astrof\'{\i}sica de Canarias (IAC), C/V\'{\i}a L\'actea sn, 38205 La Laguna, Spain
         \and
             Departamento de Astrof\'{\i}sica, Universidad de La Laguna, 38206 La Laguna, Tenerife, Spain
        \and
            Space Telescope Science Institute, Baltimore, MD, USA
         \and
            Institute of Space Sciences and Technologies of Asturias, Oviedo, Asturias, Spain
         \and
            Florida Space Institute, University of Central Florida, Orlando, FL, USA
        \and
            Observat\'orio Nacional do Rio de Janeiro, Rio de Janeiro, Brazil
        \and
            Northern Arizona University, Flagstaff, AZ, USA
        \and
            Caltech/IPAC, Pasadena, CA, USA
        \and
            Amentum, NASA Johnson Space Center, Houston, TX 77058, USA
            }

   \date{Received June 16, 2025; accepted October 03, 2025}

  \abstract
   {The analysis of the composition of primitive C-complex asteroids is essential to understand the distribution of volatiles in the Solar System since its formation. Primitive low-albedo families within the inner main asteroid belt are of particular interest because they are a significant source of carbonaceous near-Earth asteroids, such as Ryugu and Bennu.}
   {This study, part of the JWST SAMBA3 project (\textbf{S}pectral \textbf{A}nalysis of \textbf{M}ain \textbf{B}elt \textbf{A}steroids in the \textbf{3}-$\mu$m region), report the first spectroscopic analysis of asteroid (84) Klio in the 3 $\mu$m region, in order to better constrain its composition.}
   {We analysed the infrared  (0.97–5.10 $\mu$m) spectrum of Klio measured by the NIRSpec instrument on board JWST. We used the NEATM thermal model to extract the reflectance spectrum of the asteroid. Several spectral features were then analysed in the 2.8, 3.4, and 3.9 $\mu$m regions by different Gaussian fitting.}
   {Klio's spectrum shows an absorption band at 2.776 $\pm$ 0.001  $\mu$m that we attributed to phyllosilicates. We compared the position and shape of the feature with that observed in primitive materials such as carbonaceous chondrites and returned samples from Ryugu and Bennu.
   \textcolor{correc}{The position and shape of the 2.8 $\mu$m band, as well as the presence of a 0.7 $\mu$m band in the visible, suggest that Klio's spectrum is similar to certain CM2 meteorites.} We observed an absorption band around 3.9 $\mu$m, with a depth of $0.020 \pm 0.001$ that could be attributed to carbonates. We could not clearly detect any absorption associated with organics at 3.4 $\mu$m.} 
   {}

   \keywords{Methods: observational -- Techniques: spectroscopic -- Minor planets, asteroids: individual: (84) Klio
               }

   \maketitle

\section{Introduction}

\label{Sec:Intro}

Primitive C-complex asteroids are characterised by low-albedo (typically $p_V <$ 10\%) surfaces, and are commonly associated with carbonaceous chondrites (CCs), on the basis of similarities in overall spectral shapes \citep[see][and references therein]{CAMPINS2018345}. These meteorites did not reach temperatures high enough for differentiation (hence their name 'primitive'), but their initial composition still evolved, to varying degrees, through aqueous alteration or thermal metamorphism \citep{2004mete.book.....H}. CCs and primitive asteroids that have undergone aqueous alteration show evidence of hydrated minerals (such as hydrated silicates or phyllosilicates), and organic compounds \textcolor{correc}{in the form of insoluble and soluble (amino acids, aliphatic and aromatic hydrocarbons, etc.) organic matter} \citep[e.g.][]{GLAVIN2018205,2024Eleme..20...24D}. The study of primitive asteroids can reveal information about the origin and evolution of our planetary system since its early stages.

Most spectral studies of primitive asteroids have been restricted to the visible to near-infrared wavelength region (0.4-2.5 $\mu$m). Despite many observational efforts, our understanding of the composition of primitive asteroids has been limited by the paucity of diagnostic spectral features at those wavelengths, with \textcolor{correc}{few exceptions, including a} shallow absorption band around 0.7 $\mu$m, associated with Fe-rich phyllosilicates \citep{1989Sci...246..790V,2012Icar..221..744R,2014Icar..233..163F}. This band is considered a proxy of hydration: whenever it is present, an associated feature at 2.7–2.8 $\mu$m is always observed; however, the opposite is not always true \citep{2011epsc.conf..637H}. Therefore, in longer wavelengths ($>$ 2.5 $\mu$m) more diagnostic and unambiguous absorption bands are detected \citep{1995Icar..117...90R,2006Icar..182..496E, 2012Icar..219..641T,2017AJ....153...72V}. In particular, near-infrared spectroscopy up to \textcolor{correc}{4} $\mu$m enables the detection and analysis of different absorption features: a sharp band at 2.7–2.8 $\mu$m \textcolor{correc}{typically} associated with phyllosilicates \citep[e.g.][]{2012Icar..219..641T,2019PASJ...71....1U}, a rounded band around 3 $\mu$m possibly attributed to water ice \citep{2010Natur.464.1320C,2010Natur.464.1322R,2011A&A...525A..34L} or \ce{NH} groups present in ammoniated phyllosilicates \citep{2015Natur.528..241D,2022PSJ.....3..153R}, several absorption bands around 3.4 $\mu$m associated with carbon-bearing species such as organics and carbonates \citep[e.g.][]{2020Sci...370.3557K,2021A&A...653L...1K}, and absorption bands around 3.9 $\mu$m related to carbonates \citep[e.g.][]{2006Icar..185..563R}. However, the presence of strong telluric absorptions from 2.5 to 2.9 $\mu$m only allows a partial characterisation of some of these absorption bands, making compositional analysis challenging and, in some cases, impossible from ground-based telescopes.

\begin{figure}[h]
    \includegraphics[width=\columnwidth]{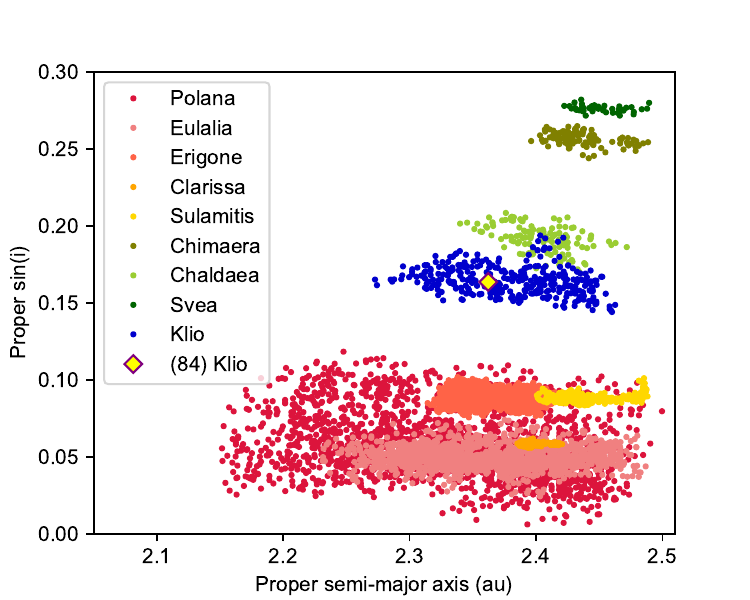}\\
    \caption{\textcolor{correc}{Proper inclination as a function of the proper semi-major axis of the low-albedo families in the inner main belt. For the Polana and Eulalia families, we used the definition from \citet{2013Icar..225..283W}, which was also used in \citet{2016Icar..266...57D}. For the other families, we applied the definition from \citet{2015aste.book..297N}. The yellow diamond marks the location of the asteroid (84) Klio.} }
    \label{fig:inclination_vs_semimajoraxis}
\end{figure}

Recently, two space missions, Hayabusa2 (JAXA) and OSIRIS-REx (NASA), returned samples from two primitive near-Earth asteroids (NEAs), Ryugu and Bennu, respectively  \citep{2022NatAs...6..214Y,2024M&PS...59.2453L}. The main goal of analysing such samples is to achieve a better and detailed understanding of the origin and evolutionary processes underwent by their parent bodies. These primitive NEAs most likely originated from primitive low-albedo collisional families located in the inner main asteroid belt, delimited by two gravitational resonances: the $\nu_6$ secular resonance with Jupiter and Saturn, at $\sim$ 2.1 au, and the 3:1 mean motion resonance with Jupiter, at $\sim$ 2.5 au \citep{2013AJ....146...26C,2015Icar..247..191B,2018Icar..313...25D}. These families include New Polana, Eulalia, Clarissa, Erigone, Sulamitis, Klio, Chaldaea, Svea, and Chimaera \textcolor{correc}{(Figures \ref{fig:inclination_vs_semimajoraxis} and \ref{fig:eccentricity_vs_semimajoraxis}). They were all} studied at visible and near-infrared wavelengths in the frame of the PRIMitive Asteroids Spectroscopic Survey (PRIMASS, \citealt{2016Icar..266...57D,2016Icar..274..231P, 2016A&A...586A.129M,2018A&A...610A..25M, 2019A&A...630A.141M, 2018Icar..311...35D, 2020A&A...643A.102D,2020Icar..33813473D,2020Icar..33513427A,2021Icar..35414028A,2021Icar..35814210A,2024Icar..41215973H}). PRIMASS is the largest dataset at these wavelengths, and at the time of writing two bundles of its associated spectral library (PRIMASS-L) have been uploaded to the NASA Planetary Data System \citep{2021pds..data....8P,2024pds..data..114P}. PRIMASS includes primitive collisional families from the entire main belt, as well as other dynamical groups, such as Hildas or Cybeles. In the present study, we focus on the asteroid (84) Klio, the largest member of the so-called Klio collisional family. In the visible range, Klio shows an absorption band at 0.7 $\mu$m (see Section \ref{Sec:Pre}), compatible with the presence of Fe-bearing phyllosilicates. This absorption band is seen in $23\%$ of all the observed Klio family members, suggesting a diversity of spectral properties within the family \citep{2019A&A...630A.141M}.\\

In this paper, we analyse the reflectance spectra of Klio using the Near-Infrared Spectrograph (NIRSpec) at the James Webb Space Telescope (JWST) \citep{2008SPIE.7010E..11C,2022A&A...661A..82B}. This object was observed as part of the SAMBA3 program (\textbf{S}pectral \textbf{A}nalysis of \textbf{M}ain \textbf{B}elt
\textbf{A}steroids in the \textbf{3}-$\mu$m region, 3$^{\text{rd}}$ General Observer Cycle, proposal \#6384) led by Driss Takir, which includes a total of nine primitive asteroids. We describe observations and data reduction, including thermal analysis in Sections \ref{Sec:Obs} and \ref{Sec:spec_term}. In Section \ref{Sec:spec} we present the spectral analysis of the detected features. We discuss the obtained results, as well as what we know about the asteroid (84) Klio itself and its collisional family from previous studies in Section \ref{Sec:discussion}. Conclusions are finally presented in Section \ref{Sec:conclusions}.

\section{Previous spectra of Klio in the visible and the near-infrared}
\label{Sec:Pre}

Asteroid Klio, as well as a number of Klio family members, have been observed in the past in the visible and near-infrared wavelengths (0.4 - 2.5 $\mu$m). Previous observations include visible spectra from SMASS-II \citep{2002Icar..158..106B}, S3OS2 \citep{2004Icar..172..179L}, and PRIMASS \citep{2019A&A...630A.141M} surveys, and near-infrared spectra from \cite{REDDY_MBASpec}, PRIMASS \citep{2020Icar..33513427A}, and D. Takir (private communication). Details on the observing conditions of these spectra are shown in Table \ref{tab:klio}. 

\begin{figure}
    \includegraphics[width=\hsize]{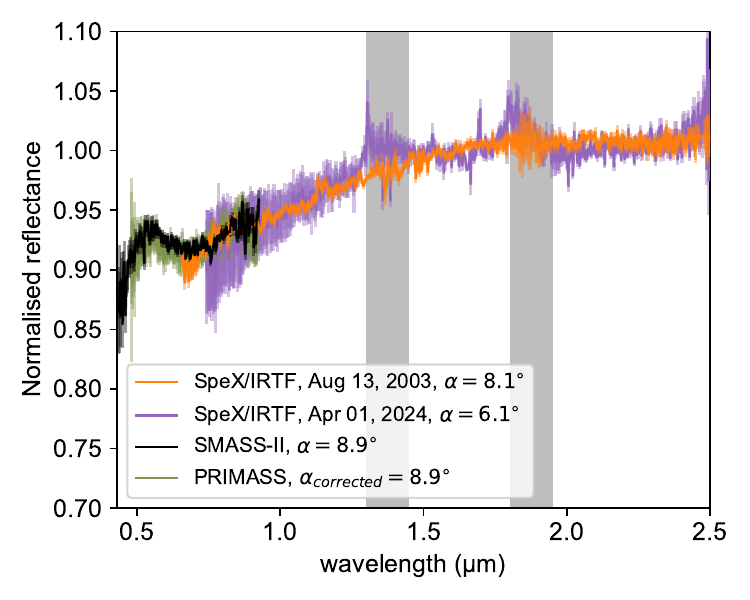}\\
    \caption{Visible and near-infrared spectra of Klio. The two spectra covering the 0.43 - 0.92 $\mu$m range are from the SMASS-II \citep[in black,][]{2002Icar..158..106B} and PRIMASS \citep[in green,][]{2024pds..data..114P} surveys, the latter after phase-reddening correction (see main text). The two spectra covering the 0.7 - 2.5 $\mu$m range were acquired by SpeX/IRTF, in 2003 \citep[in orange,][]{REDDY_MBASpec} and in 2024 (in purple). The grey areas represent the absorption by atmospheric \ce{H2O}. The complete visible to near-infrared spectra are normalised to unity at 1.6 $\mu$m.}
    \label{fig:VISNIR}
\end{figure}

We show in Fig. \ref{fig:VISNIR} the majority of the available spectra. The two near-infrared spectra obtained with the IRTF, in orange and purple, are practically identical, and were acquired at similar values of phase angle. We used the common wavelength interval, between 0.8 and 0.9 $\mu$m, to join the SMASS-II visible spectrum (in black) to the near-infrared ones. We have added here the PRIMASS visible spectrum (in black), after we corrected it from phase reddening effects. The correction was done using a slope change of approximately 0.15 \% per 1000 \r{A} per degree, as estimated for C-type asteroids by \cite{1981AJ.....86.1694L,1981AJ.....86.1705L}. This correction allows the PRIMASS visible spectrum to be rescaled to approximate the reflectance it would have exhibited if observed under the same viewing geometry as the SMASS-II observations. The two visible spectra agree very well, and both clearly show the 0.7 $\mu$m absorption band associated with Fe-bearing phyllosilicates. The S3OS2 visible spectrum is not shown here, as it is much noisier than the other two, and presents the same spectral shape. 

\begin{table}
\caption{Observational details of previous visible and near-infrared spectra of asteroid Klio.}
\label{tab:klio}
\centering
\begin{tabular}{ccccc}
\hline\hline\\[-3mm]
  Date & Telescope & Solar Analogue & $\alpha$ ($^{\circ}$) & Source\\[1mm]
\hline\\[-2mm]
\multicolumn{5}{l}{Visible spectra}\\[1mm]
\hline\\[-3mm]
06/01/94 & Hiltner & HD28099 & 8.9 & B02 \\
15/06/99 & ESO 1.52m & HD144585 & 10.9 & L04  \\
16/02/22 & GTC & SA102-1081 & 21.8 &M19 \\
\hline\\[-2mm]
\multicolumn{5}{l}{Near-infrared}\\[1mm]
\hline\\[-3mm]
13/08/13 & IRTF & HD209847 & 8.1 & R20 \\
03/02/17 & TNG & SA98-978 & 21.5 & A20 \\
01/04/24 & IRTF & SAO 157607 & 6.1 & DT \\
\hline 
\end{tabular}
\tablefoot{The table includes date, telescope, solar analogue, phase angle ($\alpha$), and data sources. Sources: B02 -- \cite[SMASS-II][]{2002Icar..158..106B}; L04 -- \cite[S3OS2][]{2004Icar..172..179L}; M19 -- \cite[PRIMASS][]{2019A&A...630A.141M}; R20 -- \cite{REDDY_MBASpec}; A20 -- \cite{2020Icar..33513427A}; DT -- Driss Takir, private communication.}
\end{table}

The near-infrared spectrum by \cite{2020Icar..33513427A} is redder than the ones shown in Fig. \ref{fig:VISNIR}, obtained with the IRTF (orange and purple). We computed the near-infrared spectral slope, $S'$, for these spectra, following the same procedure as the one defined in Arredondo et al.'s paper, and obtaining $S'$=0.483$\pm$0.020 \%/1000\AA\ for the orange one ($\alpha$=8.1$^{\circ}$) and $S'$=0.381$\pm$0.040 \%/1000\AA\ for the purple one ($\alpha$=6.1$^{\circ}$). These slopes are smaller than the one obtained by \cite{2020Icar..33513427A} for Klio, i.e $S'=1.118 \pm 0.011$ \%/1000\AA. As for the case of visible wavelengths, this difference could be attributed to differences in phase angle. There are a few references in the literature studying phase reddening in the near-infrared. Following the work of \cite{CLARK2002189} with Eros and \cite{2012Icar..220...36S} with a group of ordinary chondrites, we can use a variation in spectral slope due to phase reddening of $\sim$ 0.05 \%/1000\AA\ per degree. Applying this correction, Klio's near-infrared slope in Arredondo et al.'s paper will be $S'_{corrected}=0.448 \pm 0.011$ \%/1000\AA \ for $\alpha$=8.1$^{\circ}$, in good agreement with the slopes we obtained for the IRTF spectra. 

\section{Observations and data reduction}
\label{Sec:Obs}

The spectra of Klio presented in this paper were acquired using the integral field unit (IFU) of NIRSpec, onboard the JWST. Observations were done with the medium spectral resolution ($R \sim 1000$) grating/filter combinations: G140M/F100LP (0.97–1.84 $\mu$m), G235M/F170LP (1.66–3.07 $\mu$m), and G395M/F290LP (2.87–5.10 $\mu$m). A 4-point dither pattern was implemented for each grating setting, and the NRSIRS2RAPID readout mode was employed to optimise detector noise performance. The observations were conducted on July 6, 2024. Observational details, as well as the asteroid's physical properties retrieved from the JPL Small-Body Database Browser\footnote{https://ssd.jpl.nasa.gov/}, are shown in Table \ref{tab:properties}.

\begin{figure}
    \includegraphics[width=\columnwidth]{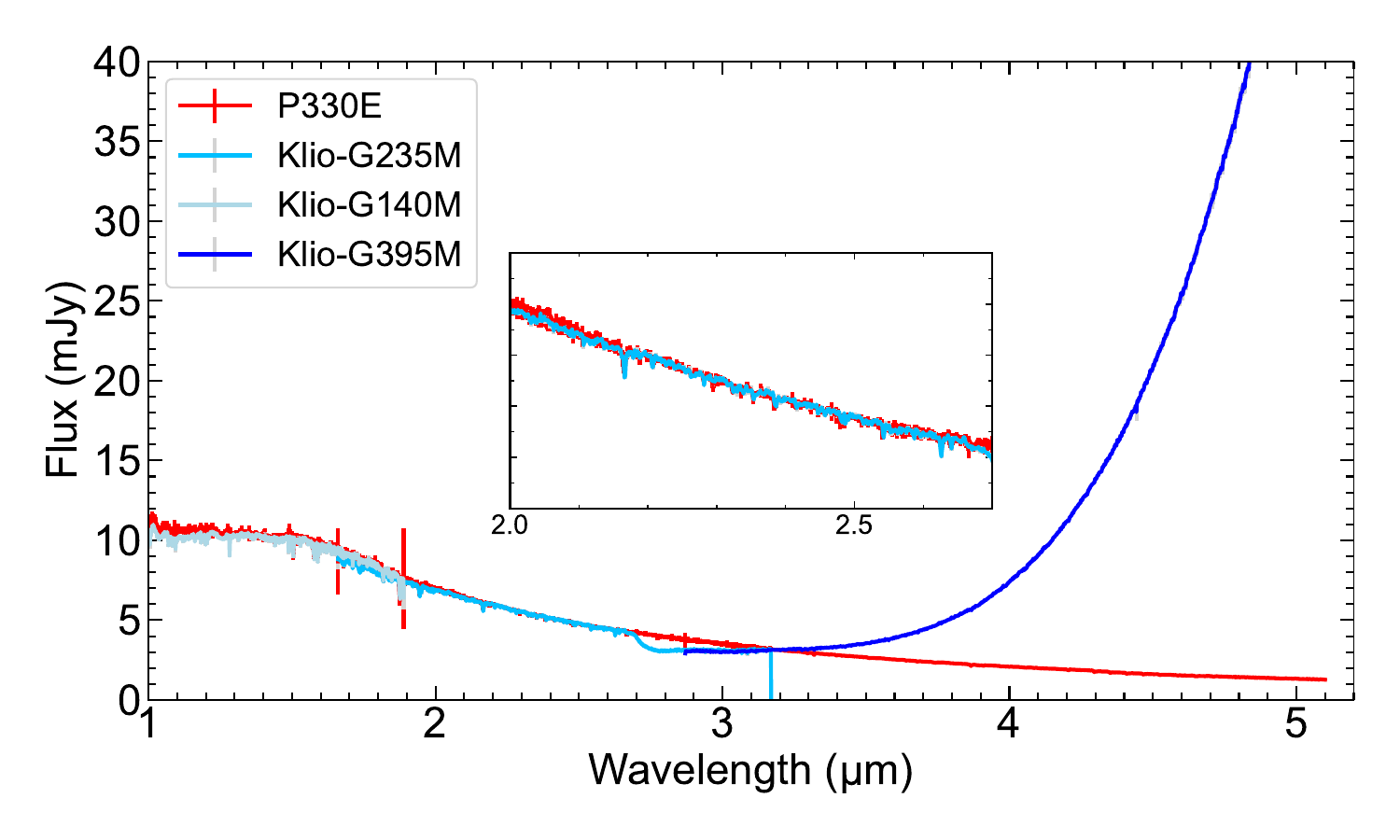}\\
    \caption{Irradiance spectrum of Klio obtained with NIRSpec using the three grating-filter combinations (in different shades of blue) shown in Table \ref{tab:properties}. The irradiance spectrum of the P330E star is shown in red, scaled to fit Klio's spectrum in the 2.2–2.6 $\mu$m region (zoomed region in the centre of the panel). Note that above $\sim$3.3 $\mu$m, Klio's spectrum becomes significantly brighter than that of P330E, indicating that thermal emission starts to dominate over reflected light.
}
    \label{fig:F1}
\end{figure}

The data processing and spectral extraction methodology was identical to that used in other recent studies of TNOs with JWST \citep[e.g.][]{Souza2024,Brunetto2025,DePra2025,Henault2025,Licandro2025,Pinilla2025}. The fully calibrated spectral data cubes, containing stacks of 2D sky-projected wavelength slices were constructed by running the raw uncalibrated data files through the first two stages of Version 1.15.1 of the official JWST pipeline \citep{Bushouse2024}, with all relevant calibration reference files drawn from context {\tt jwst\_1256.pmap} of the JWST Calibration Reference Data System. The only non-default step executed was the NSClean algorithm \citep{Rauscher2024}, which removes the 1/$f$ pattern noise in the second stage of the pipeline. A local empirical PSF was then constructed at each wavelength by median-averaging the 10 adjacent slices on either side of the considered wavelength, subtracting the background level, and normalising the template to a unit sum. Then, this PSF was fit to the original wavelength slice with a scaling factor and a constant background level. This process was applied to each dithered exposure individually, with the final combined spectrum derived by averaging each dither set together and by cleaning 3-$\sigma$ outliers on the individual spectra using a 21-pixel-wide moving median filter. The errors on the flux were estimated as the standard deviation of the four dithers. This same process was also applied to IFU observation of P330-E, a G0V calibration star \citep{Gordon2022}, using the same grating/filter combinations from PID 1538 (PI: Gordon). The irradiance spectrum of Klio is shown in Fig. \ref{fig:F1}, using different shades of blue to differentiate between the spectra obtained with each of the grating-filter combinations. We also included the  P330-E spectrum for comparison.

\begin{table*}
\caption{Observational details and asteroid's orbital and physical properties.}
\label{tab:properties}
\centering
\begin{tabular}{cccccc}
\hline\hline\\[-3mm]
  Grating/Filter & Start time (UT) & Exposure time (s) & $r$ (au) & $\Delta$ (au) & $\alpha$ ($^{\circ}$) \\[1mm]
\hline\\[-3mm]
G140M/F100LP & 12:12 & 175 &      &       &  \\
G235M/F170LP & 12:29 & 175 & 2.555 & 2.313 & 23.17 \\
G395M/F290LP & 12:47 & 175 &       &       &  \\
\hline\\[-3mm]
$a$ (au) &  $e$ &  $i$ ($^{\circ}$) & $H_V$ &  $D$ (km) & $p_V$ \\[1mm]
\hline\\[-3mm]
2.367 & 0.24 & 9.31 & 9.34 & 79.2$\pm$1.6 & 0.053$\pm$0.002\\
\hline
\end{tabular}
\tablefoot{Information includes the asteroid's distance to the Sun ($r$) and to JWST ($\Delta$), phase angle ($\alpha$), proper semi-major axis ($a$), eccentricity ($e$), inclination ($i$), absolute magnitude ($H_V$), geometric albedo ($p_V$), and diameter ($D$).}
\end{table*}

\section{Flux spectrum and thermal analysis}
\label{Sec:spec_term}

The irradiance spectrum of Klio (Fig. \ref{fig:F1}) includes reflected light and thermal emission. For thermal analysis we started carrying out a first order subtraction of the reflected light using the spectrum of the solar analogue star P330E \citep{1997AJ....113.1138C}, observed by JWST Proposal \#1538 (P.I. Karl Gordon) and extracted following the same procedure as for the case of Klio. The aperture size in the stellar spectrum extraction was chosen to match the corresponding size used for the target observation. The spectrum of P330E (in red) was first scaled to fit Klio's spectrum in the 2.2-2.6 $\mu$m region (see zoom window in Fig. \ref{fig:F1}). Notice that at wavelengths beyond $\sim$ 3.3 $\mu$m the thermal contribution on Klio's flux starts to be larger than the reflected light. As a second step, the scaled spectrum of P330E was subtracted from Klio's spectrum. The resulting almost 'pure' thermal flux spectrum in the 3.5-5.1 $\mu$m region was used to compute the diameter and geometric albedo of the asteroid, using the Near-Earth Asteroid Thermal Model \citep[NEATM,][]{1998Icar..131..291H}. 
It is known that the measured spectral energy distribution (SED) depends on the object’s size, composition, and temperature distribution. This last term is dependent on several factors, including distance to the Sun, geometric albedo, thermal inertia, surface roughness, rotation rate, shape, and spin-pole orientation. The NEATM is a refinement of the standard thermal model \citep[STM,][]{1986Icar...68..239L,1989aste.conf..128L}, which was developed and calibrated for main belt asteroids. Unlike the STM, the NEATM requires observations at multiple wavelengths and uses this information to force the model temperature distribution to be consistent with the apparent colour temperature of the asteroid. The NEATM solves simultaneously for the beaming parameter ($\eta$) and the diameter ($D$). The beaming parameter was originally introduced in the STM to allow the model temperature distribution to fit the observed enhancement of thermal emission at small solar phase angles due to surface roughness. In practice, $\eta$ can be thought of as a modelling parameter that allows a first-order correction for any effect that influences the observed surface temperature distribution (such as beaming, thermal inertia, and rotation). 

\begin{figure}
    \includegraphics[width=\columnwidth]{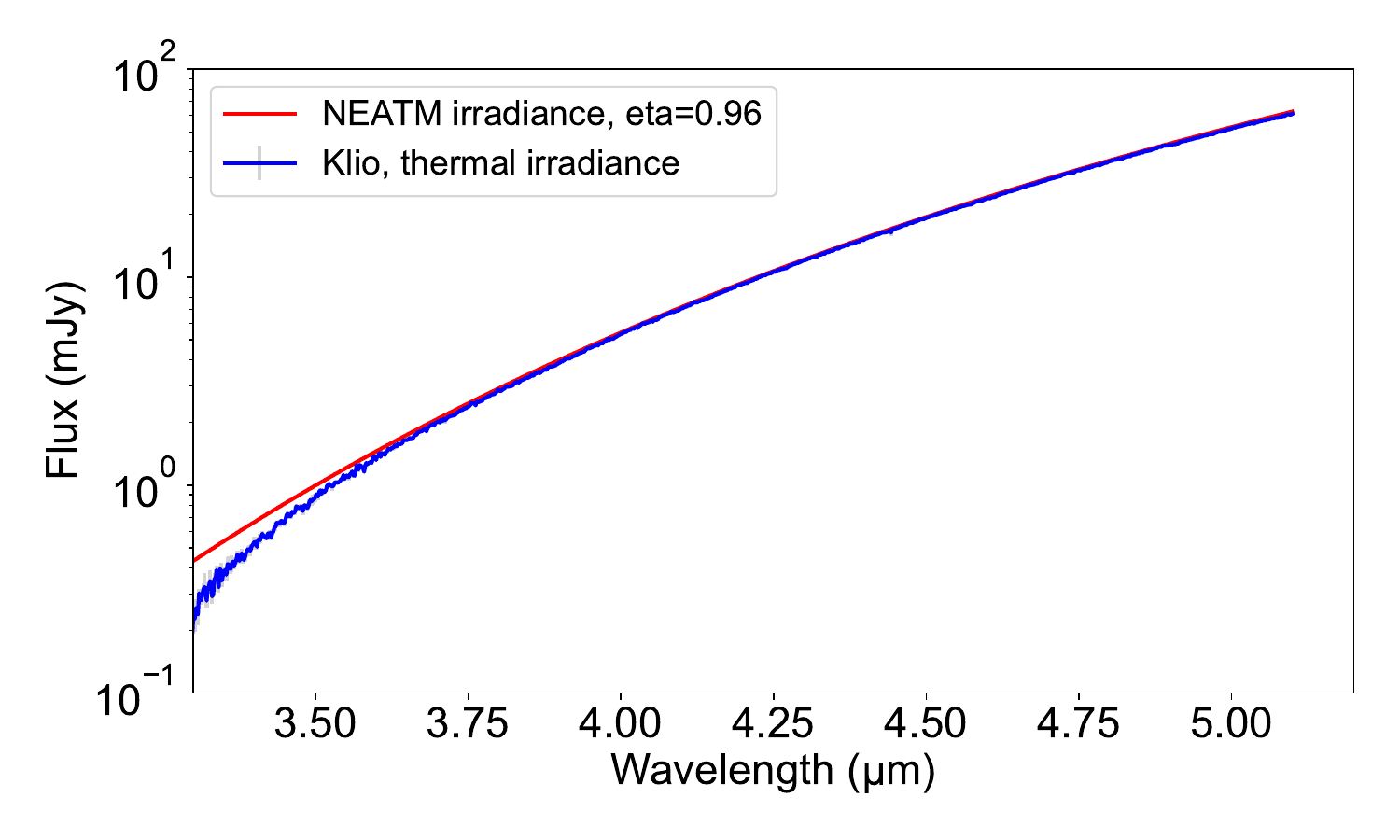}\\
    \caption{Thermal emission in Klio's spectrum (blue) fitted with the NEATM model (red). The deviation below 3.5 $\mu$m is due to the presence of a broad absorption band around 2.8 - 3 $\mu$m in Klio's spectrum.}
    \label{fig:F2}
\end{figure}

Using the NEATM (with bolometric emissivity $\epsilon$ fixed at 0.9) we fitted the thermal spectrum of Klio in the 4-5 $\mu$m region and derived a diameter $D = 78.1 \pm 23.9$ km and a beaming parameter $\eta = 0.96 \pm 0.15$. The large uncertainty in the diameter primarily arises from the limited wavelength range used for the fit, which poorly constrains the $\eta$ value. The best-fitting model is shown in Fig. \ref{fig:F2}. To estimate the visible geometric albedo $p_V$, we used the standard formula 

\begin{equation}
D = 10^{H_V/5} (1329~\textrm{km}~/ \sqrt{p_V}),
\end{equation}
with the absolute magnitude $H_V$ = 9.34 and our best fit diameter $D = 78.1$ km, obtaining a value of $p_V = 0.05 \pm 0.03$. Both the derived asteroid diameter and the computed geometric albedo are in good agreement with the ones shown in Table \ref{tab:properties}, extracted from JPL Small-Body Database Browser, obtained from IRAS Asteroid and Comet Survey observations \citep{1995yCat.7091....0V}. Likewise, these are consistent with the diameter and albedos obtained by NEOWISE \citep[$D = 82 \pm 23$ km, $p_V = 0.04 \pm 0.02$,][]{2019PDSS..251.....M}.

\section{Reflectance spectra and mineralogical analysis}
\label{Sec:spec}

\begin{figure}
    \includegraphics[width=\columnwidth]{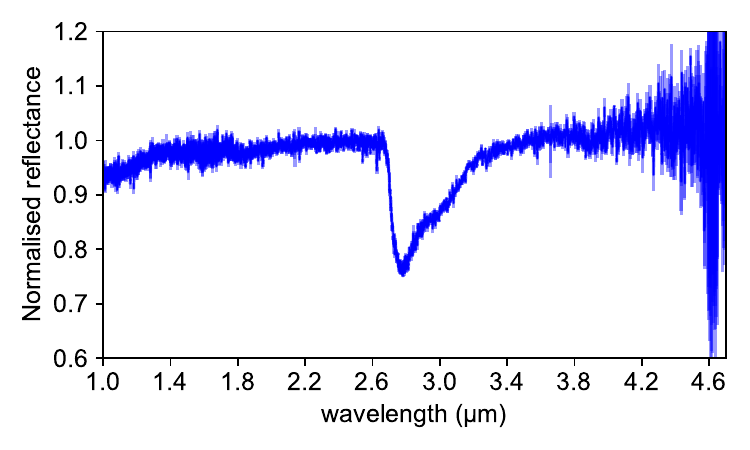}\\
    \caption{Klio reflectance spectrum. Small normalisation factors have been applied for a better fitting between the gratings, in the common wavelength region: the reflectance for grism 140M is multiplied by 0.99 and that of 395M (after thermal subtraction) is multiplied by 1.04.}
    \label{fig:F3}
\end{figure}

To derive the reflectance spectrum of the asteroid, we first subtracted the thermal emission fit obtained with the NEATM from the observed flux calibrated spectrum. Then, the resulting spectrum was divided by that of the solar analogue star, P330E. The uncertainty in the spectrum was computed as the standard deviation over the four dithers. The reflectance spectrum of Klio is shown in Fig. \ref{fig:F3}, normalised to unity at 2.6 $\mu$m for a better comparison with spectra of other small bodies obtained with JWST. The main feature observed in the spectrum is an absorption band around 2.8 $\mu$m. We can also observe hints of a minor band around 3.9 $\mu$m and an inflexion in reflectance around 3.4 $\mu$m that could be related to an absorption.

\subsection{Analysis of the 2.8 $\mu$m region}

\begin{figure}
    \includegraphics[width=\columnwidth]{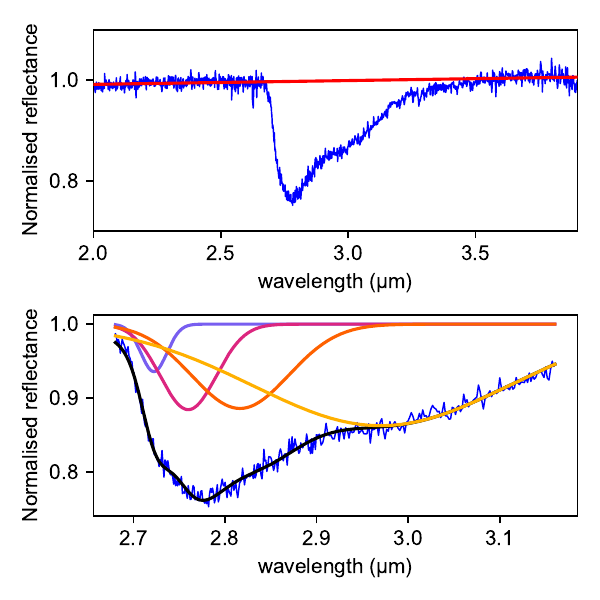}\\
    \caption{Top: linear continuum applied to the spectrum to analyse the 3 $\mu$m region. Bottom: fit of four Gaussian (in purple, pink, orange, and yellow) applied to the continuum-divided reflectance spectrum (in blue). The resulting fitting curve (sum of the Gaussian) is in black.}
    \label{fig:fit_phy}
\end{figure}

The main absorption band in Klio's spectrum is located around 2.8 $\mu$m. To analyse the position and depth of this feature, we divided the spectrum by a linear continuum, defined using two boundaries before (2.0–2.65 $\mu$m) and after (3.5–3.9 $\mu$m) the absorption band. Four Gaussian curves were then fitted to the spectrum between 2.68 and 3.16 $\mu$m (Figure \ref{fig:fit_phy}) and the band's peak position and depth were determined from the minimum of the fitted curve. To estimate the uncertainty in the spectral parameters, we resampled the spectrum for each wavelength using a uniform distribution with width equal to the respective uncertainties. This process was repeated 1,000 times to generate a distribution of fitted parameters. The final values and uncertainties were calculated as the mean and standard deviation of the resulting distribution. We obtained a peak position of $2.776 \pm 0.001$ $\mu$m and a band depth of $23.9 \pm 0.1 \%$.\\

\begin{figure*}
    \includegraphics[width=\hsize]{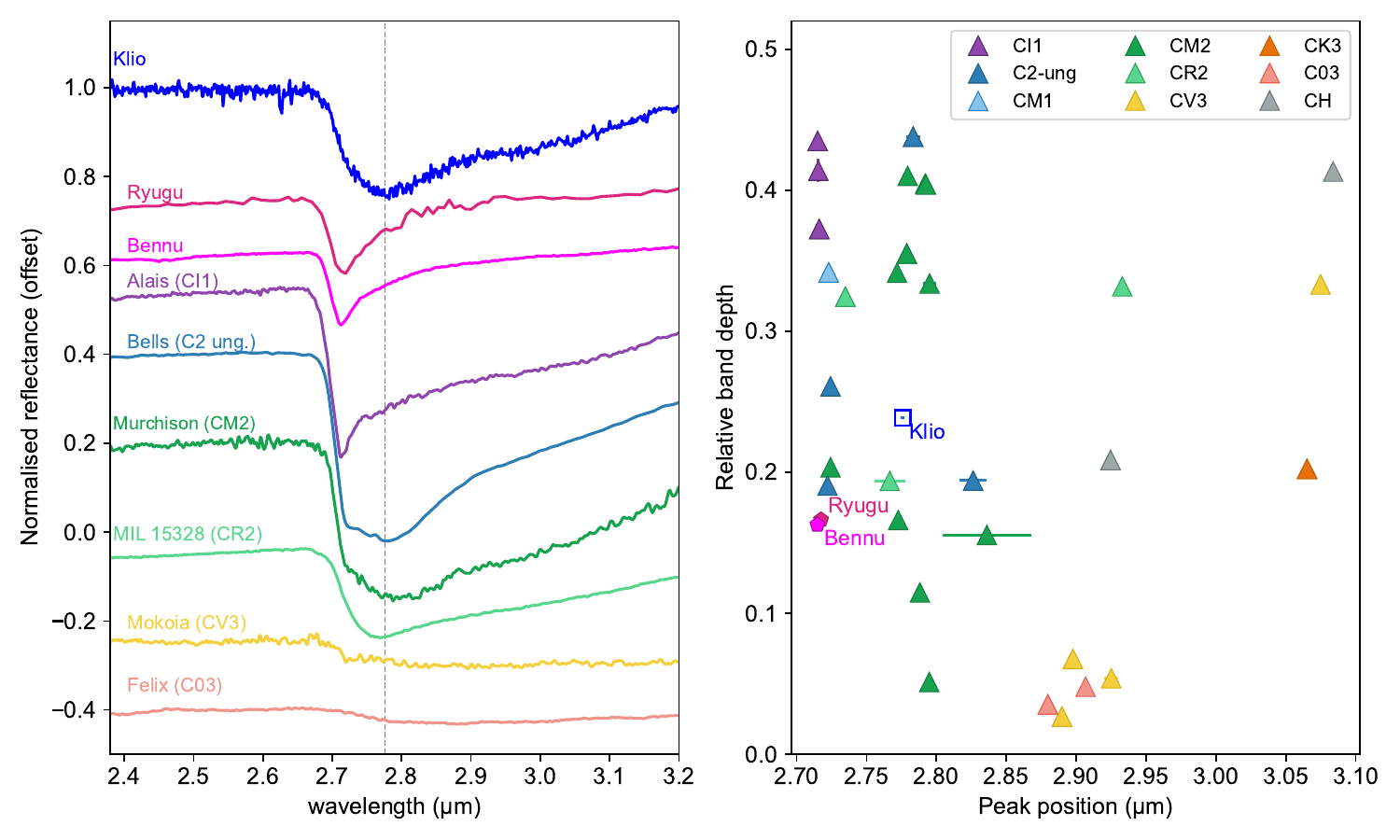}\\
     \caption{Left: Spectral comparison of the 2.8 $\mu$m band between Klio and several samples, including the spectra of CCs published in \citet{2013M&PS...48.1618T,2019Icar..333..243T}, the average spectrum of Ryugu's bulk C published in \citet{2022NatAs...6..221P}, and the average spectrum of the Bennu aggregate OREX-800029-0, published in \citet{2024M&PS...59.2453L}. The vertical dashed line indicate the position of Klio's 2.8 $\mu$m feature (i.e. 2.776 $\mu$m). An offset have been applied to the spectra for better comparison. Right: Dispersion graph between the relative band depth and the peak position of the 2.8 $\mu$m band in Klio, various CCs, Ryugu samples and Bennu samples (OREX-800029-0).}
    \label{fig:BD_pos_samples}
\end{figure*}

We compared the shape, depth, and position of Klio's 2.8 $\mu$m feature  with analogues of primitive asteroids, measured in the laboratory. These samples include several classes of CCs measured in asteroid-like conditions from \cite{2013M&PS...48.1618T,2019Icar..333..243T}, i.e., measured under vacuum and under dry conditions, in a chamber heated at temperatures up to 475 K to remove the adsorbed terrestrial water. We also compared our data with the spectra of Ryugu and Bennu samples, recently returned by the Hayabusa2 and the OSIRIS-REx missions. We used the average spectra of the two Ryugu bulk samples from chambers A and C, published in \cite{2022NatAs...6..221P}, and the average spectrum of the Bennu aggregate sample OREX-800029-0, published in \cite{2024M&PS...59.2453L}. The samples have never been exposed to Earth's atmosphere prior to the spectral measurements. The spectral comparison of Klio and the samples is shown in Fig. \ref{fig:BD_pos_samples}, left.\\

\begin{figure}
    \includegraphics[width=\columnwidth]{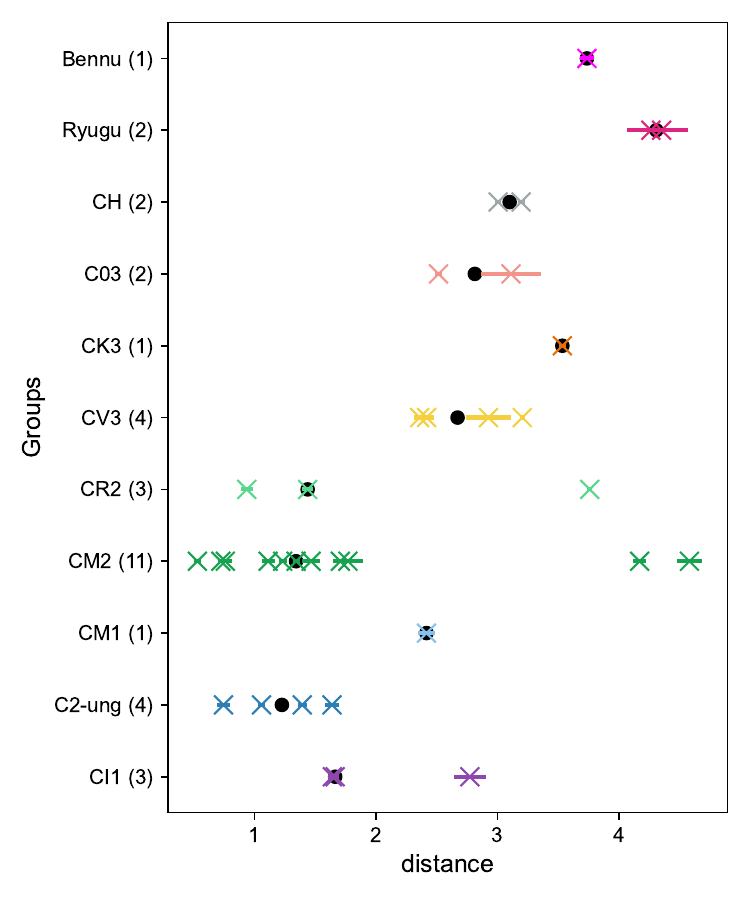}\\
    \caption{Euclidean distance of the samples (chondrites, Ryugu Bennu) to the spectrum of Klio. The samples have been merged into groups (the number of samples in each group is indicated in parenthesis). The black dot corresponds to the median of the distances within the group.}
    \label{fig:distance_euclidean}
\end{figure}

To perform a quantitative comparison, we computed the peak position and the depth of the 2.8 $\mu$m feature of all samples. Similarly to Klio spectrum, we divided the spectra of the samples by polynomial continuum across the 3 $\mu$m band before adjusting a multi-Gaussian fit to the continuum-removed spectra. Depending on the sample, we adapted the polynomial degree and the boundaries of the continuum, as well as the number of Gaussians and the boundaries of the fit. We then estimated the peak position and the depth of the band and their uncertainties using the same resampling method as for Klio's spectrum. The dispersion graph between the position and the depth of the 2.8 $\mu$m band of the samples is illustrated in Figure \ref{fig:BD_pos_samples}. To better compare Klio and the samples, we grouped the CCs samples into different classes based on their petrologic types. The peak position of Klio is generally compatible with the type 2 chondrites: C2-ungrouped, CM2 and CR2. The closest chondrites in terms of position belong to the CM2 group and are Murchison (CM2), MAC 02606 (CM2) and LAP 03786 (CM2) with peak position values of \textcolor{correc}{$2.778 \pm 0.002$, $2,773 \pm 0.004$, and $2.780 \pm 0.002$  $\mu$m}, respectively. We also compared the shape of the 2.8 $\mu$m band between Klio and the samples, by computing the spectral distance in the 2.67 - 3.10 $\mu$m range. After dividing each spectrum by its continuum, we normalised them to a band depth of 0.5 in order to emphasise on similarities in band shape rather than band depth. Then, the spectra were resampled to the same wavelength grid and the euclidean distance is calculated as follows:

\begin{equation}\label{eq:euclidean}
d = \sqrt{ \sum_{i=1}^{N} (K_{i} - S_{i})^{2} },
\end{equation}

where K and S correspond to the normalised spectra of Klio and each sample, respectively. N is the number of spectral channels in the range 2.67 - 3.10 $\mu$m. Figure \ref{fig:distance_euclidean} illustrates the euclidean distance of each sample with respect to Klio. The samples with the smallest spectral distance (i.e. the spectrum most similar to that of Klio) are LAP 03786 (CM2), QUE 97990 (CM2) and Bells (C2-ung), with distance values of $0.53 \pm 0.03 $, $0.72 \pm 0.02$, and $0.74 \pm 0.05$, respectively.

\subsection{Analysis of the 3.4 $\mu$m and 3.9 $\mu$m region}

We investigated the presence of an absorption band in the 3.4 $\mu$m region, which could be related with the presence of complex organics and/or carbonates on Klio's surface. To do this, we divided the spectrum by a 3rd order polynomial continuum, defined using two boundaries, one on each side of the 3.4 $\mu$m region: 3.1–3.3 $\mu$m and 3.7–3.8 $\mu$m. We then fitted one Gaussian curve to the spectrum to estimate its depth and position (Fig. \ref{fig:fit_orga}). As with the 2.8 $\mu$m band, we estimated uncertainties by resampling the reflectance for each wavelength within the reflectance uncertainty. After performing 1,000 Gaussian fits, we obtained a peak position of $3.492 \pm 0.130$ $\mu$m and a band depth of $0.004 \pm 0.003$. However, the position and depth of the band strongly depend on the continuum definition. For example, varying the left boundary a little, from 3.1–3.3 $\mu$m to 3.2–3.3 $\mu$m, will result in a peak position of $3.601  \pm 0.152$ $\mu$m and a band depth of $0.000 \pm 0.001$. Since the parameters depend on the continuum, and given the difficulty to clearly define the continuum because of the 3 $\mu$m band, we cannot assert with certainty that the 3.4 $\mu$m band is real.

Klio's spectrum also shows an absorption band around 3.9 $\mu$m that could indicate the presence of carbonates. We characterised it by dividing the spectrum by a 3rd order polynomial continuum defined using two boundaries, one on each side of the band: 3.6–3.7 $\mu$m and 4.03–4.4 $\mu$m range and then, performing a Gaussian fit (Fig. \ref{fig:fit_carbo}). We estimated uncertainties on the spectral parameters by performing 1,000 fits, each time resampling the reflectance for each wavelength within the uncertainty. From the fits, we obtained a peak position of $3.884 \pm 0.004$ $\mu$m, and a band depth of $0.020 \pm 0.001$.

\section{Discussion}
\label{Sec:discussion}

\subsection{Klio's composition}

In this work, to constrain the composition of Klio, we analysed the different spectral signatures of the NIRSpec data and compared them with laboratory spectra of analogue samples (CCs and returned samples). The advantage of laboratory analysis is that techniques complementary to spectroscopy can be applied to determine the composition of the samples. However, there are several limitations to the comparison between asteroids and laboratory spectra:

\begin{itemize}

    \item \textbf{Physical and observational variations}: differences in particle size and porosity as well as measurements with different \textcolor{correc}{geometries of observation (incidence, emergence and phase angles)} between asteroids and meteorites \citep[e.g.][]{CLARK2002189,BECK2012364} can affect the spectral slope, reflectance, and band depth. However, the band position remain unchanged. In this work, we compared measurements obtained with various \textcolor{correc}{geometries of observation}. Klio has been measured with a phase angle of $23^{\circ}$ while Ryugu and Bennu samples have been measured with phase angles of $35^{\circ}$ \citep[\textcolor{correc}{incidence $i\simeq35^{\circ}$, emergence $e=0^{\circ}$,}][]{2022NatAs...6..221P} and $30^{\circ}$ \citep[\textcolor{correc}{$i=30^{\circ}$, $e=0^{\circ}$,}][]{2024M&PS...59.2453L}, respectively. The spectra of the CCs published in \citet{2013M&PS...48.1618T,2019Icar..333..243T}  have \textcolor{correc}{a phase angle of 60$^{\circ}$ ($i=15^{\circ}$, $e=45^{\circ}$)}. Moreover, the surface aspect varies among the samples: the CCs are in the form of powders, while Ryugu and Bennu samples consist in coarse grains aggregates. Thus, the variations in band depth, reflectance and slope between Klio and the samples are likely affected by differences in surface aspect and observation geometry. In our analysis, we consider that the peak position is a more robust criterion than the band depth in the comparison.
    \item \textbf{Terrestrial contamination}: CCs are exposed to the terrestrial atmosphere, which can affect their composition \citep[e.g.][]{2001M&PS...36.1321G}. Meteorites absorb and adsorb water from the Earth's atmosphere and, measured by spectroscopy in ambient air, they show a very broad absorption at 3 $\mu$m, linked to \ce{H2O} molecules \citep{1994Metic..29..849M,2010GeCoA..74.4881B}. This effect has been observed when comparing Ryugu samples and CI chondrites \citep{2023SciA....9I3789A}. Here, we compared Klio with meteorites from \citet{2013M&PS...48.1618T,2019Icar..333..243T} that had been measured in a heated and under vacuum environment in order to limit the biases associated with terrestrial weathering.
    \item \textbf{Space weathering}: the surfaces of asteroids are exposed to space environment. Solar wind and cosmic rays irradiation and micrometeorite bombardment modify the physical, chemical, and spectral properties of the uppermost surface \citep[e.g.][]{2001JGR...10610039H}. Laboratory simulations of space weathering on carbonaceous chondrites have shown that the spectral slope, reflectance, band depths and band positions are affected \citep[e.g.][]{2013Icar..225..517V,2015LPI....46.1913K,2017Icar..285...43L}. We will further discuss how the position of the 2.8 $\mu$m feature in Klio's spectrum can be affected by this process.

\end{itemize}

\begin{figure}
    \includegraphics[width=\columnwidth]{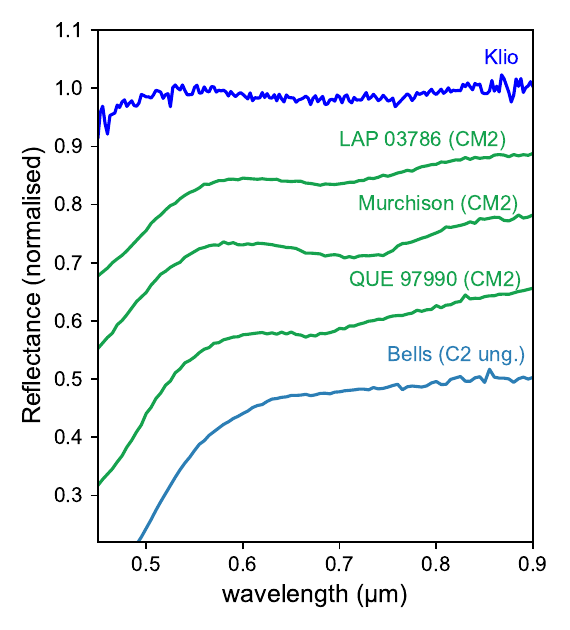}\\
    \caption{\textcolor{correc}{Spectral comparison of the 0.7 $\mu$m band between Klio \cite[SMASS-II spectrum,][]{2002Icar..158..106B} and several type 2 CCs, which are considered as good spectral match in the 3-$\mu$m region. The CCs spectra are from the RELAB database. The reference of each spectrum and sample, as well as a comparison with other CCs groups is provided in the Appendix \ref{Sec:annex_spectral_comparison_VIS}. All spectra are normalised around 0.85 $\mu$m and shifted for better comparison.}}
    \label{fig:Klio_vs_type2CC_VIS}
\end{figure}

The main feature in Klio's spectra is the absorption band located at $2.776 \pm 0.001$ $\mu$m. The position and shape of the band \textcolor{correc}{generally resemble those of type 2 CCs. In particular, the closest analogues in terms of position and shape can be found in the CM2 class (Murchison, MAC 02602, LAP 03786, QUE 97990) and the ungrouped C2 class (Bells). We can better constrain the closest spectral analogues by extending the comparison to the visible domain.  For example, the spectra of the CM2 Murchison, LAP 03786, and QUE 97990 exhibit a 0.7 $\mu$m band, which reinforces their similarity to Klio (Figure \ref{fig:Klio_vs_type2CC_VIS}). In contrast, the spectrum of Bells does not display this band. Not all CM2 meteorites have a band at 0.7 $\mu$m; for instance, the meteorites QUE 99038 and MIL 07700 do not show this feature (Figure \ref{fig:Klio_vs_alCCs_VIS}). We can also find the 0.7 $\mu$m band in CCs that do not belong to the CM2 class; this is the case for the CM1 LAP 02277 (Figure \ref{fig:Klio_vs_alCCs_VIS}) and the CR1 GRO 95577 \citep{2018Icar..313..124B}. However, for the latter, the position of the band at 2.7 $\mu$m disqualifies them as analogous to Klio. We have estimated a band position of $2.723 \pm 0.002$ $\mu$m for LAP 02277, while Beck et al. have reported a position of approximately 2.7 $\mu$m for GRO 95577. In conclusion, we can state that Klio's spectrum is most compatible with certain CM2.}

CM2 chondrites contain a significant amount of phyllosilicates (Mg-serpentine and Fe-cronstedtite), with abundances typically ranging from 56 vol.$\%$ to 84 vol.$\%$ and exhibiting varying Mg/Fe compositions \citep[e.g.][]{1997M&PS...32..231R,2009GeCoA..73.4576H,2011GeCoA..75.2735H,2015GeCoA.149..206H}. \textcolor{correc}{Variations in composition among the CM2 are partly related to the different degrees of aqueous alteration they have undergone and several scales have been established to quantify this degree \citep[e.g.][]{2007GeCoA..71.2361R,2013GeCoA.123..244A,2015GeCoA.149..206H}. In CMs, the peak position of the $\sim$ 2.7 $\mu$m band is also related to the degree of aqueous alteration. Indeed, the band is primarily attributed to the stretching vibration of OH, bound to a metallic atom in phyllosilicates \citep{10.1180/mono-4.15}. Depending on the composition and nature of the phyllosilicates, the local environment of the OH bond can vary, leading to changes in the shape and position of the band. For example, the most aqueously altered CM1 chondrites are more Mg-rich and have a peak position at shorter wavelengths compared to CM2 chondrites \citep{2010GeCoA..74.4881B,2013M&PS...48.1618T}. Linking the different scales of aqueous alteration to the position of the band in CM2 would enable a better interpretation of the degree of alteration in Klio. We have reported several alteration scales for the CM2 analysed in this paper in the appendix (Table \ref{tab:aq_alt}). However, it is challenging to precisely determine the alteration degree of Klio, as each scale probes different physicochemical properties, and there is a lack of overlap between the CM2 analysed in this work and in other studies. \citet{2007GeCoA..71.2361R} defined a subtype based on several petrologic properties, such as chondrule and matrix alteration, abundances of Fe–Ni metal, sulfides, carbonates, and opaque phases. The scale ranges from 2.0 for CM1 to 2.7–2.9 for the least aqueously altered CM so far \citep{2014GeCoA.124..190H,2014M&PS...49.1232M}. In this study, the CM2 with band positions similar to Klio (2.77–2.78 $\mu$m), corresponds to subtypes 2.1–2.5, but not all the CM2.1 have a peak position matching Klio’s. The Howard scale, which is based on the phyllosilicate / total silicate ratio, typically ranges from 1.2 to 1.7 in CM, with lower values indicating stronger alteration \citep{2015GeCoA.149..206H}. Klio’s spectral analogues fall between 1.4 and 1.5. However, there is a lack of overlap between the studies, and this comparison could only be made using seven samples in total (Table \ref{tab:aq_alt}).} Experimental simulations of solar wind irradiation by \ce{He+} ions have been conducted on CCs and phyllosilicates, and showed a shift of the 2.7 $\mu$m band towards longer wavelengths \citep{2017Icar..285...43L,2020PSJ.....1...61R} that is likely due to a decrease of the Mg/Fe ratio in phyllosilicates after irradiation. Similarly, a shift towards longer wavelength have been observed in Ryugu particles exposed to space weathering, compared to the subsurface samples \citep{2023NatAs...7.1445L,2023Icar..40615755H}. Since Klio share compositional similarities with \textcolor{correc}{some CM2}, space weathering could induce a shift of the OH band from Klio's surface towards longer wavelengths. We cannot exclude that the deeper layers of Klio have peak positions at shorter wavelengths compared to the surface, pointing to Fe-bearing phyllosilicates with a slightly more Mg-rich composition.\\

In primitive asteroids, the 3.4 $\mu$m region is relevant to detect features associated with organics or carbonates \citep[e.g.][]{2010Natur.464.1320C,2010Natur.464.1322R,2011A&A...525A..34L,2020Sci...370.3557K}, and the 3.9 $\mu$m region is also characteristic of carbonates \citep[e.g.][]{2006Icar..185..563R}. We detected an absorption band around $3.9$ $\mu$m, with a band depth of  $0.020 \pm 0.001$ that could be attributed to the $\nu_1 + \nu_3$ combination mode of the \ce{CO3^2-} vibration in carbonates \citep[e.g.][]{2021E&SS....801844B,2024E&SS...1103666B}. On the other side, we were unable to definitively identify the presence of the carbonate and/or organic band around 3.4 $\mu$m with our method. The feature could be weak because space weathering would have partly degraded the organic molecules on Klio's surface. The weakening of the 3.4 $\mu$m band has been observed in an irradiation experiment of a Ceres simulant containing undecanoic acid (\ce{C10H21COOH}) \citep{DESANCTIS2024}.

\subsection{Relationship with other low-albedo asteroids}

\textcolor{correc}{Both the 2.8 and the 0.7 $\mu$m bands} indicate that Klio's hydrated minerals could have a composition similar to \textcolor{correc}{certain CM2}, a class that contains Mg- and Fe-rich phyllosilicates. In the visible range, Klio family members show spectral heterogeneity. \citet{2019A&A...630A.141M} studied 30 asteroids belonging to Klio's family and showed that $23\%$ of them exhibit the 0.7 $\mu$m feature. The $23\%$ of asteroids showing the band is expected to have more iron-rich phyllosilicates than the others. The remaining $77\%$ could either have fewer phyllosilicates in general or have phyllosilicates that are richer in Mg. Understanding which of these two cases is correct, or whether both are, is an open question but important for understanding the degree of aqueous alteration of the parent body of the Klio family.\\

Previous studies dismissed the Klio family as a potential origin for Ryugu and Bennu, due to spectral differences \citep{2013AJ....146...26C,2020Icar..33513427A}, and because dynamical models and observations favour low-inclination families such as New Polana \citep{2013AJ....146...26C,2015Icar..247..191B}. In this work, we show that the 2.8 $\mu$m band of Klio has a distinct position and shape compared to the samples from the two asteroids. This result reinforces the hypothesis that Klio is an unlikely parent body for the Hayabusa2 and OSIRIS-REx targets. Analyses of Ryugu and Bennu samples have shown that their composition is more similar to CI1 chondrites \citep{2022NatAs...6.1163I,2023Sci...379.8671N,2023NatAs...7..398Y,2024M&PS...59.2453L} whereas our results suggest that Klio is more similar to \textcolor{correc}{some CM2.}

\section{Conclusions}
\label{Sec:conclusions}

We analysed Klio's spectrum obtained with NIRSpec onboard JWST. We used the NEATM thermal model to derive a geometric albedo of $p_V = 0.05 \pm 0.03$, and a diameter of $D = 78.1 \pm 23.9$. The thermal analysis enabled us to isolate the reflected flux from the thermal flux. We then analysed the reflectance spectrum of Klio and identified several absorption bands related to the aqueous alteration history of the asteroid. In particular, we identified a feature at $2.776 \pm 0.001$ $\mu$m, related to phyllosilicates. The shape and the position of the band suggest that Klio could have a composition similar to some \textcolor{correc}{specific CM2 meteorites}. We could not identify features associated with organics but we observed a band at 3.9 $\mu$m, suggesting the presence of carbonates on Klio's surface.

\begin{acknowledgements}

We would like to acknowledge the support of the Space Telescope Science Institute (JWST-GO-06384.001-A). JL, JdL, TlP acknowledge support from the Agencia Estatal de Investigacion del Ministerio de Ciencia e Innovaci\'on (AEI-MCINN) under grant 'Hydrated Minerals and Organic Compounds in Primitive Asteroids'
 with reference PID2020-120464GB-100. The authors thank Dante Lauretta and Cédric Pilorget, for providing the spectra of Bennu and Ryugu bulk samples.
\textcolor{correc}{This research utilises spectra acquired by K. T. Howard, T. Hiroi, M. E. Zolensky, M. D. Dyar, and J. M. Sunshine  with the NASA RELAB facility at Brown University. We would like to thank the referee, Pierre Beck, for his constructive comments.}

\end{acknowledgements}

\bibliographystyle{aa}
\bibliography{Klio_JWSTRef}

\begin{thebibliography}{99}
\expandafter\ifx\csname natexlab\endcsname\relax\def\natexlab#1{#1}\fi

\bibitem[{{Alexander} {et~al.}(2013){Alexander}, {Howard}, {Bowden}, \&
  {Fogel}}]{2013GeCoA.123..244A}
{Alexander}, C. M.~O.~D., {Howard}, K.~T., {Bowden}, R., \& {Fogel}, M.~L.
  2013, \gca, 123, 244

\bibitem[{{Amano} {et~al.}(2023){Amano}, {Matsuoka}, {Nakamura}, {Kagawa},
  {Fujioka}, {Potin}, {Hiroi}, {Tatsumi}, {Milliken}, {Quirico}, {Beck},
  {Brunetto}, {Uesugi}, {Takahashi}, {Kawai}, {Yamashita}, {Enokido}, {Wada},
  {Furukawa}, {Zolensky}, {Takir}, {Domingue}, {Jaramillo-Correa}, {Vilas},
  {Hendrix}, {Kikuiri}, {Morita}, {Yurimoto}, {Noguchi}, {Okazaki}, {Yabuta},
  {Naraoka}, {Sakamoto}, {Tachibana}, {Yada}, {Nishimura}, {Nakato},
  {Miyazaki}, {Yogata}, {Abe}, {Okada}, {Usui}, {Yoshikawa}, {Saiki}, {Tanaka},
  {Terui}, {Nakazawa}, {Watanabe}, \& {Tsuda}}]{2023SciA....9I3789A}
{Amano}, K., {Matsuoka}, M., {Nakamura}, T., {et~al.} 2023, Science Advances,
  9, eadi3789

\bibitem[{{Arredondo} {et~al.}(2021{\natexlab{a}}){Arredondo}, {Campins},
  {Pinilla-Alonso}, {de Le{\'o}n}, {Lorenzi}, \&
  {Morate}}]{2021Icar..35414028A}
{Arredondo}, A., {Campins}, H., {Pinilla-Alonso}, N., {et~al.}
  2021{\natexlab{a}}, \icarus, 354, 114028

\bibitem[{{Arredondo} {et~al.}(2021{\natexlab{b}}){Arredondo}, {Campins},
  {Pinilla-Alonso}, {de Le{\'o}n}, {Lorenzi}, \&
  {Morate}}]{2021Icar..35814210A}
{Arredondo}, A., {Campins}, H., {Pinilla-Alonso}, N., {et~al.}
  2021{\natexlab{b}}, \icarus, 358, 114210

\bibitem[{{Arredondo} {et~al.}(2020){Arredondo}, {Lorenzi}, {Pinilla-Alonso},
  {Campins}, {Malfavon}, {de Le{\'o}n}, \& {Morate}}]{2020Icar..33513427A}
{Arredondo}, A., {Lorenzi}, V., {Pinilla-Alonso}, N., {et~al.} 2020, \icarus,
  335, 113427

\bibitem[{{Beck} {et~al.}(2024){Beck}, {Beyssac}, {Schmitt}, {Royer}, {Mandon},
  {Boulard}, {Rividi}, \& {Cloutis}}]{2024E&SS...1103666B}
{Beck}, P., {Beyssac}, O., {Schmitt}, B., {et~al.} 2024, Earth and Space
  Science, 11, e2024EA003666

\bibitem[{{Beck} {et~al.}(2018){Beck}, {Maturilli}, {Garenne}, {Vernazza},
  {Helbert}, {Quirico}, \& {Schmitt}}]{2018Icar..313..124B}
{Beck}, P., {Maturilli}, A., {Garenne}, A., {et~al.} 2018, \icarus, 313, 124

\bibitem[{Beck {et~al.}(2012)Beck, Pommerol, Thomas, Schmitt, Moynier, \&
  Barrat}]{BECK2012364}
Beck, P., Pommerol, A., Thomas, N., {et~al.} 2012, Icarus, 218, 364

\bibitem[{{Beck} {et~al.}(2010){Beck}, {Quirico}, {Montes-Hernandez}, {Bonal},
  {Bollard}, {Orthous-Daunay}, {Howard}, {Schmitt}, {Brissaud}, {Deschamps},
  {Wunder}, \& {Guillot}}]{2010GeCoA..74.4881B}
{Beck}, P., {Quirico}, E., {Montes-Hernandez}, G., {et~al.} 2010, \gca, 74,
  4881

\bibitem[{{Bishop} {et~al.}(2021){Bishop}, {King}, {Lane}, {Brown}, {Lafuente},
  {Hiroi}, {Roberts}, {Swayze}, {Lin}, \& {S{\'a}nchez
  Rom{\'a}n}}]{2021E&SS....801844B}
{Bishop}, J.~L., {King}, S.~J., {Lane}, M.~D., {et~al.} 2021, Earth and Space
  Science, 8, e01844

\bibitem[{{B{\"o}ker} {et~al.}(2022){B{\"o}ker}, {Arribas}, {L{\"u}tzgendorf},
  {Alves de Oliveira}, {Beck}, {Birkmann}, {Bunker}, {Charlot}, {de Marchi},
  {Ferruit}, {Giardino}, {Jakobsen}, {Kumari}, {L{\'o}pez-Caniego}, {Maiolino},
  {Manjavacas}, {Marston}, {Moseley}, {Muzerolle}, {Ogle}, {Pirzkal},
  {Rauscher}, {Rawle}, {Rix}, {Sabbi}, {Sargent}, {Sirianni}, {te Plate},
  {Valenti}, {Willott}, \& {Zeidler}}]{2022A&A...661A..82B}
{B{\"o}ker}, T., {Arribas}, S., {L{\"u}tzgendorf}, N., {et~al.} 2022, \aap,
  661, A82

\bibitem[{{Bottke} {et~al.}(2015){Bottke}, {Vokrouhlick{\'y}}, {Walsh},
  {Delbo}, {Michel}, {Lauretta}, {Campins}, {Connolly}, {Scheeres}, \&
  {Chelsey}}]{2015Icar..247..191B}
{Bottke}, W.~F., {Vokrouhlick{\'y}}, D., {Walsh}, K.~J., {et~al.} 2015,
  \icarus, 247, 191

\bibitem[{{Brunetto} {et~al.}(2025){Brunetto}, {H{\'e}nault}, {Cryan},
  {Pinilla-Alonso}, {Emery}, {Guilbert-Lepoutre}, {Holler}, {McClure},
  {M{\"u}ller}, {Pendleton}, {de Souza-Feliciano}, {Stansberry}, {Grundy},
  {Peixinho}, {Strazzulla}, {Bannister}, {Cruikshank}, {Harvison}, {Licandro},
  {Lorenzi}, {de Pr{\'a}}, \& {Schambeau}}]{Brunetto2025}
{Brunetto}, R., {H{\'e}nault}, E., {Cryan}, S., {et~al.} 2025, \apjl, 982, L8

\bibitem[{{Bus} \& {Binzel}(2002)}]{2002Icar..158..106B}
{Bus}, S.~J. \& {Binzel}, R.~P. 2002, \icarus, 158, 106

\bibitem[{{Bushouse} {et~al.}(2024){Bushouse}, {Eisenhamer}, {Dencheva},
  {Davies}, {Greenfield}, {Morrison}, {Hodge}, {Simon}, {Grumm}, {Droettboom},
  {Slavich}, {Sosey}, {Pauly}, {Miller}, {Jedrzejewski}, {Hack}, {Davis},
  {Crawford}, {Law}, {Gordon}, {Regan}, {Cara}, {MacDonald}, {Bradley},
  {Shanahan}, {Jamieson}, {Teodoro}, {Williams}, \&
  {Pena-Guerrero}}]{Bushouse2024}
{Bushouse}, H., {Eisenhamer}, J., {Dencheva}, N., {et~al.} 2024, {JWST
  Calibration Pipeline}

\bibitem[{{Campins} {et~al.}(2013){Campins}, {de Le{\'o}n}, {Morbidelli},
  {Licandro}, {Gayon-Markt}, {Delbo}, \& {Michel}}]{2013AJ....146...26C}
{Campins}, H., {de Le{\'o}n}, J., {Morbidelli}, A., {et~al.} 2013, \aj, 146, 26

\bibitem[{Campins {et~al.}(2018)Campins, de~León, Licandro, Hendrix, Sánchez,
  \& Ali-Lagoa}]{CAMPINS2018345}
Campins, H., de~León, J., Licandro, J., {et~al.} 2018, in Primitive Meteorites
  and Asteroids, ed. N.~Abreu (Elsevier), 345--369

\bibitem[{{Campins} {et~al.}(2010){Campins}, {Hargrove}, {Pinilla-Alonso},
  {Howell}, {Kelley}, {Licandro}, {Moth{\'e}-Diniz}, {Fern{\'a}ndez}, \&
  {Ziffer}}]{2010Natur.464.1320C}
{Campins}, H., {Hargrove}, K., {Pinilla-Alonso}, N., {et~al.} 2010, \nat, 464,
  1320

\bibitem[{Clark {et~al.}(2002)Clark, Helfenstein, Bell, Peterson, Veverka,
  Izenberg, Domingue, Wellnitz, \& McFadden}]{CLARK2002189}
Clark, B.~E., Helfenstein, P., Bell, J., {et~al.} 2002, Icarus, 155, 189

\bibitem[{{Closs} {et~al.}(2008){Closs}, {Ferruit}, {Lobb}, {Preuss}, {Rolt},
  \& {Talbot}}]{2008SPIE.7010E..11C}
{Closs}, M.~F., {Ferruit}, P., {Lobb}, D.~R., {et~al.} 2008, in Society of
  Photo-Optical Instrumentation Engineers (SPIE) Conference Series, Vol. 7010,
  Space Telescopes and Instrumentation 2008: Optical, Infrared, and Millimeter,
  ed. J.~{Oschmann}, Jacobus~M., M.~W.~M. {de Graauw}, \& H.~A. {MacEwen},
  701011

\bibitem[{{Colina} \& {Bohlin}(1997)}]{1997AJ....113.1138C}
{Colina}, L. \& {Bohlin}, R. 1997, \aj, 113, 1138

\bibitem[{{Davidson} {et~al.}(2019){Davidson}, {Alexander}, {King}, {Bates},
  {Foustoukos}, {Schrader}, {Bullock}, {Busemann}, {Riebe},
  {Sch{\"o}nb{\"a}chler}, \& {Clay}}]{2019LPICo2189.2115D}
{Davidson}, J., {Alexander}, C.~M.~O., {King}, A.~J., {et~al.} 2019, in LPI
  Contributions, Vol. 2189, Asteroid Science in the Age of Hayabusa2 and
  OSIRIS-REx, 2115

\bibitem[{{De Gregorio} \& {Engrand}(2024)}]{2024Eleme..20...24D}
{De Gregorio}, B.~T. \& {Engrand}, C. 2024, Elements, 20, 24

\bibitem[{{de Le{\'o}n} {et~al.}(2018){de Le{\'o}n}, {Campins}, {Morate}, {De
  Pr{\'a}}, {Al{\'\i}-Lagoa}, {Licandro}, {Rizos}, {Pinilla-Alonso},
  {DellaGiustina}, {Lauretta}, {Popescu}, \& {Lorenzi}}]{2018Icar..313...25D}
{de Le{\'o}n}, J., {Campins}, H., {Morate}, D., {et~al.} 2018, \icarus, 313, 25

\bibitem[{{de Le{\'o}n} {et~al.}(2016){de Le{\'o}n}, {Pinilla-Alonso}, {Delbo},
  {Campins}, {Cabrera-Lavers}, {Tanga}, {Cellino}, {Bendjoya}, {Gayon-Markt},
  {Licandro}, {Lorenzi}, {Morate}, {Walsh}, {DeMeo}, {Landsman}, \&
  {Al{\'\i}-Lagoa}}]{2016Icar..266...57D}
{de Le{\'o}n}, J., {Pinilla-Alonso}, N., {Delbo}, M., {et~al.} 2016, \icarus,
  266, 57

\bibitem[{{De Pr{\'a}} {et~al.}(2025){De Pr{\'a}}, {H{\'e}nault},
  {Pinilla-Alonso}, {Holler}, {Brunetto}, {Stansberry}, {de Souza Feliciano},
  {Carvano}, {Harvison}, {Licandro}, {M{\"u}ller}, {Peixinho}, {Lorenzi},
  {Guilbert-Lepoutre}, {Bannister}, {Pendleton}, {Cruikshank}, {Schambeau},
  {McClure}, \& {Emery}}]{DePra2025}
{De Pr{\'a}}, M.~N., {H{\'e}nault}, E., {Pinilla-Alonso}, N., {et~al.} 2025,
  Nature Astronomy, 9, 252

\bibitem[{{De Pr{\'a}} {et~al.}(2020{\natexlab{a}}){De Pr{\'a}}, {Licandro},
  {Pinilla-Alonso}, {Lorenzi}, {Rond{\'o}n}, {Carvano}, {Morate}, \& {De
  Le{\'o}n}}]{2020Icar..33813473D}
{De Pr{\'a}}, M.~N., {Licandro}, J., {Pinilla-Alonso}, N., {et~al.}
  2020{\natexlab{a}}, \icarus, 338, 113473

\bibitem[{{De Pr{\'a}} {et~al.}(2020{\natexlab{b}}){De Pr{\'a}},
  {Pinilla-Alonso}, {Carvano}, {Licandro}, {Morate}, {Lorenzi}, {de Le{\'o}n},
  {Campins}, \& {Moth{\'e}-Diniz}}]{2020A&A...643A.102D}
{De Pr{\'a}}, M.~N., {Pinilla-Alonso}, N., {Carvano}, J., {et~al.}
  2020{\natexlab{b}}, \aap, 643, A102

\bibitem[{{De Pr{\'a}} {et~al.}(2018){De Pr{\'a}}, {Pinilla-Alonso}, {Carvano},
  {Licandro}, {Campins}, {Moth{\'e}-Diniz}, {De Le{\'o}n}, \&
  {Al{\'\i}-Lagoa}}]{2018Icar..311...35D}
{De Pr{\'a}}, M.~N., {Pinilla-Alonso}, N., {Carvano}, J.~M., {et~al.} 2018,
  \icarus, 311, 35

\bibitem[{{De Sanctis} {et~al.}(2015){De Sanctis}, {Ammannito}, {Raponi},
  {Marchi}, {McCord}, {McSween}, {Capaccioni}, {Capria}, {Carrozzo},
  {Ciarniello}, {Longobardo}, {Tosi}, {Fonte}, {Formisano}, {Frigeri},
  {Giardino}, {Magni}, {Palomba}, {Turrini}, {Zambon}, {Combe}, {Feldman},
  {Jaumann}, {McFadden}, {Pieters}, {Prettyman}, {Toplis}, {Raymond}, \&
  {Russell}}]{2015Natur.528..241D}
{De Sanctis}, M.~C., {Ammannito}, E., {Raponi}, A., {et~al.} 2015, \nat, 528,
  241

\bibitem[{{De Sanctis} {et~al.}(2024){De Sanctis}, Baratta, Brucato,
  Castillo-Rogez, Ciarniello, Cozzolino, Angelis, Ferrari, Fulvio, Germanà,
  Mennella, Pagnoscin, Palumbo, Poggiali, Popa, Raponi, Scirè, Strazzulla, \&
  Urso}]{DESANCTIS2024}
{De Sanctis}, M.~C., Baratta, G.~A., Brucato, J.~R., {et~al.} 2024, Science
  Advances, 10, eadp3664

\bibitem[{{Emery} {et~al.}(2006){Emery}, {Cruikshank}, \& {Van
  Cleve}}]{2006Icar..182..496E}
{Emery}, J.~P., {Cruikshank}, D.~P., \& {Van Cleve}, J. 2006, \icarus, 182, 496

\bibitem[{Farmer(1974)}]{10.1180/mono-4.15}
Farmer, V.~C. 1974, in The Infrared Spectra of Minerals (Mineralogical Society
  of Great Britain and Ireland)

\bibitem[{{Fornasier} {et~al.}(2014){Fornasier}, {Lantz}, {Barucci}, \&
  {Lazzarin}}]{2014Icar..233..163F}
{Fornasier}, S., {Lantz}, C., {Barucci}, M.~A., \& {Lazzarin}, M. 2014,
  \icarus, 233, 163

\bibitem[{Glavin {et~al.}(2018)Glavin, Alexander, Aponte, Dworkin, Elsila, \&
  Yabuta}]{GLAVIN2018205}
Glavin, D.~P., Alexander, C.~M., Aponte, J.~C., {et~al.} 2018, in Primitive
  Meteorites and Asteroids, ed. N.~Abreu (Elsevier), 205--271

\bibitem[{{Gordon} {et~al.}(2022){Gordon}, {Bohlin}, {Sloan}, {Rieke}, {Volk},
  {Boyer}, {Muzerolle}, {Schlawin}, {Deustua}, {Hines}, {Kraemer}, {Mullally},
  \& {Su}}]{Gordon2022}
{Gordon}, K.~D., {Bohlin}, R., {Sloan}, G.~C., {et~al.} 2022, \aj, 163, 267

\bibitem[{{Gounelle} \& {Zolensky}(2001)}]{2001M&PS...36.1321G}
{Gounelle}, M. \& {Zolensky}, M.~E. 2001, \maps, 36, 1321

\bibitem[{{Hapke}(2001)}]{2001JGR...10610039H}
{Hapke}, B. 2001, \jgr, 106, 10039

\bibitem[{{Harris}(1998)}]{1998Icar..131..291H}
{Harris}, A.~W. 1998, \icarus, 131, 291

\bibitem[{{Harvison} {et~al.}(2024){Harvison}, {De Pr{\'a}}, {Pinilla-Alonso},
  {Lorenzi}, {de Le{\'o}n}, {Morate}, {Licandro}, {Arredondo}, \&
  {Campins}}]{2024Icar..41215973H}
{Harvison}, B., {De Pr{\'a}}, M., {Pinilla-Alonso}, N., {et~al.} 2024, \icarus,
  412, 115973

\bibitem[{{H{\'e}nault} {et~al.}(2025){H{\'e}nault}, {Brunetto},
  {Pinilla-Alonso}, {Baklouti}, {Djouadi}, {Guilbert-Lepoutre}, {M{\"u}ller},
  {Cryan}, {de Souza-Feliciano}, {Holler}, {de Pr{\'a}}, {Emery}, {McClure},
  {Schambeau}, {Pendleton}, {Harvison}, {Licandro}, {Lorenzi}, {Cruikshank},
  {Peixinho}, {Bannister}, \& {Stansberry}}]{Henault2025}
{H{\'e}nault}, E., {Brunetto}, R., {Pinilla-Alonso}, N., {et~al.} 2025, \aap,
  694, A126

\bibitem[{{Hewins} {et~al.}(2014){Hewins}, {Bourot-Denise}, {Zanda}, {Leroux},
  {Barrat}, {Humayun}, {G{\"o}pel}, {Greenwood}, {Franchi}, {Pont}, {Lorand},
  {Courn{\`e}de}, {Gattacceca}, {Rochette}, {Kuga}, {Marrocchi}, \&
  {Marty}}]{2014GeCoA.124..190H}
{Hewins}, R.~H., {Bourot-Denise}, M., {Zanda}, B., {et~al.} 2014, \gca, 124,
  190

\bibitem[{{Hiroi} {et~al.}(2023){Hiroi}, {Milliken}, {Robertson}, {Schultz},
  {Amano}, {Nakamura}, {Yurimoto}, {Noguchi}, {Okazaki}, {Naraoka}, {Yabuta},
  {Sakamoto}, {Yada}, {Nishimura}, {Nakato}, {Miyazaki}, {Yogata}, {Abe},
  {Okada}, {Usui}, {Yoshikawa}, {Saiki}, {Tanaka}, {Nakazawa}, {Yokota},
  {Tatsumi}, {Tsuda}, {Tachibana}, {Fuyuto}, {Watanabe}, {Sasaki}, {Kaiden},
  {Kitazato}, \& {Matsuoka}}]{2023Icar..40615755H}
{Hiroi}, T., {Milliken}, R.~E., {Robertson}, K.~M., {et~al.} 2023, \icarus,
  406, 115755

\bibitem[{{Howard} {et~al.}(2015){Howard}, {Alexander}, {Schrader}, \&
  {Dyl}}]{2015GeCoA.149..206H}
{Howard}, K.~T., {Alexander}, C.~M.~O.~D., {Schrader}, D.~L., \& {Dyl}, K.~A.
  2015, \gca, 149, 206

\bibitem[{{Howard} {et~al.}(2009){Howard}, {Benedix}, {Bland}, \&
  {Cressey}}]{2009GeCoA..73.4576H}
{Howard}, K.~T., {Benedix}, G.~K., {Bland}, P.~A., \& {Cressey}, G. 2009, \gca,
  73, 4576

\bibitem[{{Howard} {et~al.}(2011){Howard}, {Benedix}, {Bland}, \&
  {Cressey}}]{2011GeCoA..75.2735H}
{Howard}, K.~T., {Benedix}, G.~K., {Bland}, P.~A., \& {Cressey}, G. 2011, \gca,
  75, 2735

\bibitem[{{Howell} {et~al.}(2011){Howell}, {Rivkin}, {Vilas}, {Magri}, {Nolan},
  {Vervack}, \& {Fernandez}}]{2011epsc.conf..637H}
{Howell}, E.~S., {Rivkin}, A.~S., {Vilas}, F., {et~al.} 2011, in EPSC-DPS Joint
  Meeting 2011, Vol. 2011, 637

\bibitem[{{Hutchison}(2004)}]{2004mete.book.....H}
{Hutchison}, R. 2004, {Meteorites} (Cambridge University Press)

\bibitem[{{Ito} {et~al.}(2022){Ito}, {Tomioka}, {Uesugi}, {Yamaguchi},
  {Shirai}, {Ohigashi}, {Liu}, {Greenwood}, {Kimura}, {Imae}, {Uesugi},
  {Nakato}, {Yogata}, {Yuzawa}, {Kodama}, {Tsuchiyama}, {Yasutake}, {Findlay},
  {Franchi}, {Malley}, {McCain}, {Matsuda}, {McKeegan}, {Hirahara}, {Takeuchi},
  {Sekimoto}, {Sakurai}, {Okada}, {Karouji}, {Arakawa}, {Fujii}, {Fujimoto},
  {Hayakawa}, {Hirata}, {Hirata}, {Honda}, {Honda}, {Hosoda}, {Iijima},
  {Ikeda}, {Ishiguro}, {Ishihara}, {Iwata}, {Kawahara}, {Kikuchi}, {Kitazato},
  {Matsumoto}, {Matsuoka}, {Michikami}, {Mimasu}, {Miura}, {Mori}, {Morota},
  {Nakazawa}, {Namiki}, {Noda}, {Noguchi}, {Ogawa}, {Ogawa}, {Okada},
  {Okamoto}, {Ono}, {Ozaki}, {Saiki}, {Sakatani}, {Sawada}, {Senshu},
  {Shimaki}, {Shirai}, {Sugita}, {Takei}, {Takeuchi}, {Tanaka}, {Tatsumi},
  {Terui}, {Tsukizaki}, {Wada}, {Yamada}, {Yamada}, {Yamamoto}, {Yano},
  {Yokota}, {Yoshihara}, {Yoshikawa}, {Yoshikawa}, {Fukai}, {Furuya},
  {Hatakeda}, {Hayashi}, {Hitomi}, {Kumagai}, {Miyazaki}, {Nishimura},
  {Soejima}, {Iwamae}, {Yamamoto}, {Yoshitake}, {Yada}, {Abe}, {Usui},
  {Watanabe}, \& {Tsuda}}]{2022NatAs...6.1163I}
{Ito}, M., {Tomioka}, N., {Uesugi}, M., {et~al.} 2022, Nature Astronomy, 6,
  1163

\bibitem[{{Kaplan} {et~al.}(2020){Kaplan}, {Lauretta}, {Simon}, {Hamilton},
  {DellaGiustina}, {Golish}, {Reuter}, {Bennett}, {Burke}, {Campins},
  {Connolly}, {Dworkin}, {Emery}, {Glavin}, {Glotch}, {Hanna}, {Ishimaru},
  {Jawin}, {McCoy}, {Porter}, {Sandford}, {Ferrone}, {Clark}, {Li}, {Zou},
  {Daly}, {Barnouin}, {Seabrook}, \& {Enos}}]{2020Sci...370.3557K}
{Kaplan}, H.~H., {Lauretta}, D.~S., {Simon}, A.~A., {et~al.} 2020, Science,
  370, eabc3557

\bibitem[{{Kaplan} {et~al.}(2021){Kaplan}, {Simon}, {Hamilton}, {Thompson},
  {Sandford}, {Barucci}, {Cloutis}, {Brucato}, {Reuter}, {Glavin}, {Clark},
  {Dworkin}, {Campins}, {Emery}, {Fornasier}, {Zou}, \&
  {Lauretta}}]{2021A&A...653L...1K}
{Kaplan}, H.~H., {Simon}, A.~A., {Hamilton}, V.~E., {et~al.} 2021, \aap, 653,
  L1

\bibitem[{{Keller} {et~al.}(2015){Keller}, {Christoffersen}, {Dukes},
  {Baragiola}, \& {Rahman}}]{2015LPI....46.1913K}
{Keller}, L.~P., {Christoffersen}, R., {Dukes}, C.~A., {Baragiola}, R., \&
  {Rahman}, Z. 2015, in 46th Annual Lunar and Planetary Science Conference,
  Lunar and Planetary Science Conference, 1913

\bibitem[{{Krietsch} {et~al.}(2021){Krietsch}, {Busemann}, {Riebe}, {King},
  {Alexander}, \& {Maden}}]{2021GeCoA.310..240K}
{Krietsch}, D., {Busemann}, H., {Riebe}, M. E.~I., {et~al.} 2021, \gca, 310,
  240

\bibitem[{{Lantz} {et~al.}(2017){Lantz}, {Brunetto}, {Barucci}, {Fornasier},
  {Baklouti}, {Bour{\c{c}}ois}, \& {Godard}}]{2017Icar..285...43L}
{Lantz}, C., {Brunetto}, R., {Barucci}, M.~A., {et~al.} 2017, \icarus, 285, 43

\bibitem[{{Lauretta} {et~al.}(2024){Lauretta}, {Connolly}, {Aebersold},
  {Alexander}, {Ballouz}, {Barnes}, {Bates}, {Bennett}, {Blanche},
  {Blumenfeld}, {Clemett}, {Cody}, {DellaGiustina}, {Dworkin}, {Eckley},
  {Foustoukos}, {Franchi}, {Glavin}, {Greenwood}, {Haenecour}, {Hamilton},
  {Hill}, {Hiroi}, {Ishimaru}, {Jourdan}, {Kaplan}, {Keller}, {King},
  {Koefoed}, {Kontogiannis}, {Le}, {Macke}, {McCoy}, {Milliken}, {Najorka},
  {Nguyen}, {Pajola}, {Polit}, {Righter}, {Roper}, {Russell}, {Ryan},
  {Sandford}, {Schofield}, {Schultz}, {Seifert}, {Tachibana}, {Thomas-Keprta},
  {Thompson}, {Tu}, {Tusberti}, {Wang}, {Zega}, \&
  {Wolner}}]{2024M&PS...59.2453L}
{Lauretta}, D.~S., {Connolly}, H.~C., {Aebersold}, J.~E., {et~al.} 2024, \maps,
  59, 2453

\bibitem[{{Lazzaro} {et~al.}(2004){Lazzaro}, {Angeli}, {Carvano},
  {Moth{\'e}-Diniz}, {Duffard}, \& {Florczak}}]{2004Icar..172..179L}
{Lazzaro}, D., {Angeli}, C.~A., {Carvano}, J.~M., {et~al.} 2004, \icarus, 172,
  179

\bibitem[{{Le Pivert-Jolivet} {et~al.}(2023){Le Pivert-Jolivet}, {Brunetto},
  {Pilorget}, {Bibring}, {Nakato}, {Hamm}, {Hatakeda}, {Lantz}, {Loizeau},
  {Riu}, {Yogata}, {Baklouti}, {Poulet}, {Al{\'e}on-Toppani}, {Carter},
  {Langevin}, {Okada}, {Yada}, {Hitomi}, {Kumagai}, {Miyazaki}, {Nagashima},
  {Nishimura}, {Usui}, {Abe}, {Saiki}, {Tanaka}, {Nakazawa}, {Tsuda}, \&
  {Watanabe}}]{2023NatAs...7.1445L}
{Le Pivert-Jolivet}, T., {Brunetto}, R., {Pilorget}, C., {et~al.} 2023, Nature
  Astronomy, 7, 1445

\bibitem[{{Lebofsky} \& {Spencer}(1989)}]{1989aste.conf..128L}
{Lebofsky}, L.~A. \& {Spencer}, J.~R. 1989, in Asteroids II, ed. R.~P.
  {Binzel}, T.~{Gehrels}, \& M.~S. {Matthews}, 128--147

\bibitem[{{Lebofsky} {et~al.}(1986){Lebofsky}, {Sykes}, {Tedesco}, {Veeder},
  {Matson}, {Brown}, {Gradie}, {Feierberg}, \& {Rudy}}]{1986Icar...68..239L}
{Lebofsky}, L.~A., {Sykes}, M.~V., {Tedesco}, E.~F., {et~al.} 1986, \icarus,
  68, 239

\bibitem[{{Licandro} {et~al.}(2011){Licandro}, {Campins}, {Kelley}, {Hargrove},
  {Pinilla-Alonso}, {Cruikshank}, {Rivkin}, \& {Emery}}]{2011A&A...525A..34L}
{Licandro}, J., {Campins}, H., {Kelley}, M., {et~al.} 2011, \aap, 525, A34

\bibitem[{{Licandro} {et~al.}(2025){Licandro}, {Pinilla-Alonso}, {Holler}, {De
  Pr{\'a}}, {Melita}, {de Souza Feliciano}, {Brunetto}, {Guilbert-Lepoutre},
  {H{\'e}nault}, {Lorenzi}, {Stansberry}, {Schambeau}, {Harvison}, {Pendleton},
  {Cruikshank}, {M{\"u}ller}, {McClure}, {Emery}, {Peixinho}, {Bannister}, \&
  {Wong}}]{Licandro2025}
{Licandro}, J., {Pinilla-Alonso}, N., {Holler}, B.~J., {et~al.} 2025, Nature
  Astronomy, 9, 245

\bibitem[{{Lumme} \& {Bowell}(1981{\natexlab{a}})}]{1981AJ.....86.1694L}
{Lumme}, K. \& {Bowell}, E. 1981{\natexlab{a}}, \aj, 86, 1694

\bibitem[{{Lumme} \& {Bowell}(1981{\natexlab{b}})}]{1981AJ.....86.1705L}
{Lumme}, K. \& {Bowell}, E. 1981{\natexlab{b}}, \aj, 86, 1705

\bibitem[{{Mainzer} {et~al.}(2019){Mainzer}, {Bauer}, {Cutri}, {Grav},
  {Kramer}, {Masiero}, {Sonnett}, \& {Wright}}]{2019PDSS..251.....M}
{Mainzer}, A.~K., {Bauer}, J.~M., {Cutri}, R.~M., {et~al.} 2019, {NEOWISE
  Diameters and Albedos V2.0}, NASA Planetary Data System,
  urn:nasa:pds:neowise\_diameters\_albedos::2.0

\bibitem[{{Marrocchi} {et~al.}(2014){Marrocchi}, {Gounelle}, {Blanchard},
  {Caste}, \& {Kearsley}}]{2014M&PS...49.1232M}
{Marrocchi}, Y., {Gounelle}, M., {Blanchard}, I., {Caste}, F., \& {Kearsley},
  A.~T. 2014, \maps, 49, 1232

\bibitem[{{Miyamoto} \& {Zolensky}(1994)}]{1994Metic..29..849M}
{Miyamoto}, M. \& {Zolensky}, M.~E. 1994, Meteoritics, 29, 849

\bibitem[{{Morate} {et~al.}(2018){Morate}, {de Le{\'o}n}, {De Pr{\'a}},
  {Licandro}, {Cabrera-Lavers}, {Campins}, \&
  {Pinilla-Alonso}}]{2018A&A...610A..25M}
{Morate}, D., {de Le{\'o}n}, J., {De Pr{\'a}}, M., {et~al.} 2018, \aap, 610,
  A25

\bibitem[{{Morate} {et~al.}(2016){Morate}, {de Le{\'o}n}, {De Pr{\'a}},
  {Licandro}, {Cabrera-Lavers}, {Campins}, {Pinilla-Alonso}, \&
  {Al{\'\i}-Lagoa}}]{2016A&A...586A.129M}
{Morate}, D., {de Le{\'o}n}, J., {De Pr{\'a}}, M., {et~al.} 2016, \aap, 586,
  A129

\bibitem[{{Morate} {et~al.}(2019){Morate}, {de Le{\'o}n}, {De Pr{\'a}},
  {Licandro}, {Pinilla-Alonso}, {Campins}, {Arredondo}, {Carvano}, {Lazzaro},
  \& {Cabrera-Lavers}}]{2019A&A...630A.141M}
{Morate}, D., {de Le{\'o}n}, J., {De Pr{\'a}}, M., {et~al.} 2019, \aap, 630,
  A141

\bibitem[{{Nakamura} {et~al.}(2023){Nakamura}, {Matsumoto}, {Amano}, {Enokido},
  {Zolensky}, {Mikouchi}, {Genda}, {Tanaka}, {Zolotov}, {Kurosawa}, {Wakita},
  {Hyodo}, {Nagano}, {Nakashima}, {Takahashi}, {Fujioka}, {Kikuiri}, {Kagawa},
  {Matsuoka}, {Brearley}, {Tsuchiyama}, {Uesugi}, {Matsuno}, {Kimura}, {Sato},
  {Milliken}, {Tatsumi}, {Sugita}, {Hiroi}, {Kitazato}, {Brownlee}, {Joswiak},
  {Takahashi}, {Ninomiya}, {Takahashi}, {Osawa}, {Terada}, {Brenker},
  {Tkalcec}, {Vincze}, {Brunetto}, {Al{\'e}on-Toppani}, {Chan}, {Roskosz},
  {Viennet}, {Beck}, {Alp}, {Michikami}, {Nagaashi}, {Tsuji}, {Ino},
  {Martinez}, {Han}, {Dolocan}, {Bodnar}, {Tanaka}, {Yoshida}, {Sugiyama},
  {King}, {Fukushi}, {Suga}, {Yamashita}, {Kawai}, {Inoue}, {Nakato},
  {Noguchi}, {Vilas}, {Hendrix}, {Jaramillo-Correa}, {Domingue}, {Dominguez},
  {Gainsforth}, {Engrand}, {Duprat}, {Russell}, {Bonato}, {Ma}, {Kawamoto},
  {Wada}, {Watanabe}, {Endo}, {Enju}, {Riu}, {Rubino}, {Tack}, {Takeshita},
  {Takeichi}, {Takeuchi}, {Takigawa}, {Takir}, {Tanigaki}, {Taniguchi},
  {Tsukamoto}, {Yagi}, {Yamada}, {Yamamoto}, {Yamashita}, {Yasutake}, {Uesugi},
  {Umegaki}, {Chiu}, {Ishizaki}, {Okumura}, {Palomba}, {Pilorget}, {Potin},
  {Alasli}, {Anada}, {Araki}, {Sakatani}, {Schultz}, {Sekizawa}, {Sitzman},
  {Sugiura}, {Sun}, {Dartois}, {De Pauw}, {Dionnet}, {Djouadi}, {Falkenberg},
  {Fujita}, {Fukuma}, {Gearba}, {Hagiya}, {Hu}, {Kato}, {Kawamura}, {Kimura},
  {Kubo}, {Langenhorst}, {Lantz}, {Lavina}, {Lindner}, {Zhao}, {Vekemans},
  {Baklouti}, {Bazi}, {Borondics}, {Nagasawa}, {Nishiyama}, {Nitta},
  {Mathurin}, {Matsumoto}, {Mitsukawa}, {Miura}, {Miyake}, {Miyake},
  {Yurimoto}, {Okazaki}, {Yabuta}, {Naraoka}, {Sakamoto}, {Tachibana},
  {Connolly}, {Lauretta}, {Yoshitake}, {Yoshikawa}, {Yoshikawa}, {Yoshihara},
  {Yokota}, {Yogata}, {Yano}, {Yamamoto}, {Yamamoto}, {Yamada}, {Yamada},
  {Yada}, {Wada}, {Usui}, {Tsukizaki}, {Terui}, {Takeuchi}, {Takei}, {Iwamae},
  {Soejima}, {Shirai}, {Shimaki}, {Senshu}, {Sawada}, {Saiki}, {Ozaki}, {Ono},
  {Okada}, {Ogawa}, {Ogawa}, {Noguchi}, {Noda}, {Nishimura}, {Namiki},
  {Nakazawa}, {Morota}, {Miyazaki}, {Miura}, {Mimasu}, {Matsumoto}, {Kumagai},
  {Kouyama}, {Kikuchi}, {Kawahara}, \& {Kameda}}]{2023Sci...379.8671N}
{Nakamura}, T., {Matsumoto}, M., {Amano}, K., {et~al.} 2023, Science, 379,
  abn8671

\bibitem[{{Nesvorn{\'y}} {et~al.}(2015){Nesvorn{\'y}}, {Bro{\v{z}}}, \&
  {Carruba}}]{2015aste.book..297N}
{Nesvorn{\'y}}, D., {Bro{\v{z}}}, M., \& {Carruba}, V. 2015, in Asteroids IV,
  ed. P.~{Michel}, F.~E. {DeMeo}, \& W.~F. {Bottke}, 297--321

\bibitem[{{Pilorget} {et~al.}(2021){Pilorget}, {Okada}, {Hamm}, {Brunetto},
  {Yada}, {Loizeau}, {Riu}, {Usui}, {Moussi-Soffys}, {Hatakeda}, {Nakato},
  {Yogata}, {Abe}, {Al{\'e}on-Toppani}, {Carter}, {Chaigneau}, {Crane},
  {Gondet}, {Kumagai}, {Langevin}, {Lantz}, {Le Pivert-Jolivet}, {Lequertier},
  {Lourit}, {Miyazaki}, {Nishimura}, {Poulet}, {Arakawa}, {Hirata}, {Kitazato},
  {Nakazawa}, {Namiki}, {Saiki}, {Sugita}, {Tachibana}, {Tanaka}, {Yoshikawa},
  {Tsuda}, {Watanabe}, \& {Bibring}}]{2022NatAs...6..221P}
{Pilorget}, C., {Okada}, T., {Hamm}, V., {et~al.} 2021, Nature Astronomy, 6,
  221

\bibitem[{{Pinilla-Alonso} {et~al.}(2025){Pinilla-Alonso}, {Brunetto}, {De
  Pr{\'a}}, {Holler}, {H{\'e}nault}, {Feliciano}, {Lorenzi}, {Pendleton},
  {Cruikshank}, {M{\"u}ller}, {Stansberry}, {Emery}, {Schambeau}, {Licandro},
  {Harvison}, {McClure}, {Guilbert-Lepoutre}, {Peixinho}, {Bannister}, \&
  {Wong}}]{Pinilla2025}
{Pinilla-Alonso}, N., {Brunetto}, R., {De Pr{\'a}}, M.~N., {et~al.} 2025,
  Nature Astronomy, 9, 230

\bibitem[{{Pinilla-Alonso} {et~al.}(2016){Pinilla-Alonso}, {de Le{\'o}n},
  {Walsh}, {Campins}, {Lorenzi}, {Delbo}, {DeMeo}, {Licandro}, {Landsman},
  {Lucas}, {Al{\'\i}-Lagoa}, \& {Burt}}]{2016Icar..274..231P}
{Pinilla-Alonso}, N., {de Le{\'o}n}, J., {Walsh}, K.~J., {et~al.} 2016,
  \icarus, 274, 231

\bibitem[{{Pinilla-Alonso} {et~al.}(2021){Pinilla-Alonso}, {De Pra}, {de Leon},
  {Morate}, {Lorenzi}, {Arredondo}, {Campins}, {Licandro}, {Delbo},
  {Cabrera-Lavers}, {Walsh}, {DeMeo}, \& {Sarid}}]{2021pds..data....8P}
{Pinilla-Alonso}, N., {De Pra}, M., {de Leon}, J., {et~al.} 2021, {PRIMASS-L
  Spectra Bundle V1.0}, NASA Planetary Data System,
  urn:nasa:pds:gbo.ast.primass-l.spectra::1.0

\bibitem[{{Pinilla-Alonso} {et~al.}(2024){Pinilla-Alonso}, {De Pra}, {de
  Le{\'o}n}, {Harvison}, {Morate}, {Lorenzi}, {Arredondo}, {Campins},
  {Licandro}, {Delbo}, {Cabrera-Lavers}, {Walsh}, {DeMeo}, \&
  {Sarid}}]{2024pds..data..114P}
{Pinilla-Alonso}, N., {De Pra}, M.~N., {de Le{\'o}n}, J., {et~al.} 2024,
  {PRIMASS-L Spectra Bundle V2.0}, NASA Planetary Data System,
  urn:nasa:pds:gbo.ast.primass-l.spectra::2.0

\bibitem[{{Rauscher}(2024)}]{Rauscher2024}
{Rauscher}, B.~J. 2024, \pasp, 136, 015001

\bibitem[{Reddy \& Sanchez(2020)}]{REDDY_MBASpec}
Reddy, V. \& Sanchez, J.~A. 2020, Reddy Main Belt Asteroid Spectra V1.0,
  urn:nasa:pds:gbo.ast-mb.reddy.spectra::1.0

\bibitem[{{Rivkin}(2012)}]{2012Icar..221..744R}
{Rivkin}, A.~S. 2012, \icarus, 221, 744

\bibitem[{{Rivkin} \& {Emery}(2010)}]{2010Natur.464.1322R}
{Rivkin}, A.~S. \& {Emery}, J.~P. 2010, \nat, 464, 1322

\bibitem[{{Rivkin} {et~al.}(2022){Rivkin}, {Emery}, {Howell}, {Kareta},
  {Noonan}, {Richardson}, {Sharkey}, {Sickafoose}, {Woodney}, {Cartwright},
  {Lindsay}, \& {Mcclure}}]{2022PSJ.....3..153R}
{Rivkin}, A.~S., {Emery}, J.~P., {Howell}, E.~S., {et~al.} 2022, \psj, 3, 153

\bibitem[{{Rivkin} {et~al.}(1995){Rivkin}, {Howell}, {Britt}, {Lebofsky},
  {Nolan}, \& {Branston}}]{1995Icar..117...90R}
{Rivkin}, A.~S., {Howell}, E.~S., {Britt}, D.~T., {et~al.} 1995, \icarus, 117,
  90

\bibitem[{{Rivkin} {et~al.}(2006){Rivkin}, {Volquardsen}, \&
  {Clark}}]{2006Icar..185..563R}
{Rivkin}, A.~S., {Volquardsen}, E.~L., \& {Clark}, B.~E. 2006, \icarus, 185,
  563

\bibitem[{{Rubin}(1997)}]{1997M&PS...32..231R}
{Rubin}, A.~E. 1997, \maps, 32, 231

\bibitem[{{Rubin} {et~al.}(2007){Rubin}, {Trigo-Rodr{\'\i}guez}, {Huber}, \&
  {Wasson}}]{2007GeCoA..71.2361R}
{Rubin}, A.~E., {Trigo-Rodr{\'\i}guez}, J.~M., {Huber}, H., \& {Wasson}, J.~T.
  2007, \gca, 71, 2361

\bibitem[{{Rubino} {et~al.}(2020){Rubino}, {Lantz}, {Baklouti}, {Leroux},
  {Borondics}, \& {Brunetto}}]{2020PSJ.....1...61R}
{Rubino}, S., {Lantz}, C., {Baklouti}, D., {et~al.} 2020, \psj, 1, 61

\bibitem[{{Sanchez} {et~al.}(2012){Sanchez}, {Reddy}, {Nathues}, {Cloutis},
  {Mann}, \& {Hiesinger}}]{2012Icar..220...36S}
{Sanchez}, J.~A., {Reddy}, V., {Nathues}, A., {et~al.} 2012, \icarus, 220, 36

\bibitem[{{Souza-Feliciano} {et~al.}(2024){Souza-Feliciano}, {Holler},
  {Pinilla-Alonso}, {De Pr{\'a}}, {Brunetto}, {M{\"u}ller}, {Stansberry},
  {Licandro}, {Emery}, {Henault}, {Guilbert-Lepoutre}, {Pendleton},
  {Cruikshank}, {Schambeau}, {Bannister}, {Peixinho}, {McClure}, {Harvison}, \&
  {Lorenzi}}]{Souza2024}
{Souza-Feliciano}, A.~C., {Holler}, B.~J., {Pinilla-Alonso}, N., {et~al.} 2024,
  \aap, 681, L17

\bibitem[{{Takir} \& {Emery}(2012)}]{2012Icar..219..641T}
{Takir}, D. \& {Emery}, J.~P. 2012, \icarus, 219, 641

\bibitem[{{Takir} {et~al.}(2013){Takir}, {Emery}, {McSween}, {Hibbitts},
  {Clark}, {Pearson}, \& {Wang}}]{2013M&PS...48.1618T}
{Takir}, D., {Emery}, J.~P., {McSween}, H.~Y., {et~al.} 2013, \maps, 48, 1618

\bibitem[{{Takir} {et~al.}(2019){Takir}, {Stockstill-Cahill}, {Hibbitts}, \&
  {Nakauchi}}]{2019Icar..333..243T}
{Takir}, D., {Stockstill-Cahill}, K.~R., {Hibbitts}, C.~A., \& {Nakauchi}, Y.
  2019, \icarus, 333, 243

\bibitem[{{Usui} {et~al.}(2019){Usui}, {Hasegawa}, {Ootsubo}, \&
  {Onaka}}]{2019PASJ...71....1U}
{Usui}, F., {Hasegawa}, S., {Ootsubo}, T., \& {Onaka}, T. 2019, \pasj, 71, 1

\bibitem[{{Veeder} \& {Walker}(1995)}]{1995yCat.7091....0V}
{Veeder}, G.~J. \& {Walker}, R.~G. 1995, {VizieR Online Data Catalog: IRAS
  Asteroid and Comet Survey (Veeder+ 1986)}, VizieR On-line Data Catalog:
  VII/91. Originally published in: IRAS Asteroid and Comet Survey, IPAC, JPL
  D-3698, 1986

\bibitem[{{Vernazza} {et~al.}(2017){Vernazza}, {Castillo-Rogez}, {Beck},
  {Emery}, {Brunetto}, {Delbo}, {Marsset}, {Marchis}, {Groussin}, {Zanda},
  {Lamy}, {Jorda}, {Mousis}, {Delsanti}, {Djouadi}, {Dionnet}, {Borondics}, \&
  {Carry}}]{2017AJ....153...72V}
{Vernazza}, P., {Castillo-Rogez}, J., {Beck}, P., {et~al.} 2017, \aj, 153, 72

\bibitem[{{Vernazza} {et~al.}(2013){Vernazza}, {Fulvio}, {Brunetto}, {Emery},
  {Dukes}, {Cipriani}, {Witasse}, {Schaible}, {Zanda}, {Strazzulla}, \&
  {Baragiola}}]{2013Icar..225..517V}
{Vernazza}, P., {Fulvio}, D., {Brunetto}, R., {et~al.} 2013, \icarus, 225, 517

\bibitem[{{Vilas} \& {Gaffey}(1989)}]{1989Sci...246..790V}
{Vilas}, F. \& {Gaffey}, M.~J. 1989, Science, 246, 790

\bibitem[{{Walsh} {et~al.}(2013){Walsh}, {Delb{\'o}}, {Bottke},
  {Vokrouhlick{\'y}}, \& {Lauretta}}]{2013Icar..225..283W}
{Walsh}, K.~J., {Delb{\'o}}, M., {Bottke}, W.~F., {Vokrouhlick{\'y}}, D., \&
  {Lauretta}, D.~S. 2013, \icarus, 225, 283

\bibitem[{{Yada} {et~al.}(2021){Yada}, {Abe}, {Okada}, {Nakato}, {Yogata},
  {Miyazaki}, {Hatakeda}, {Kumagai}, {Nishimura}, {Hitomi}, {Soejima},
  {Yoshitake}, {Iwamae}, {Furuya}, {Uesugi}, {Karouji}, {Usui}, {Hayashi},
  {Yamamoto}, {Fukai}, {Sugita}, {Cho}, {Yumoto}, {Yabe}, {Bibring},
  {Pilorget}, {Hamm}, {Brunetto}, {Riu}, {Lourit}, {Loizeau}, {Lequertier},
  {Moussi-Soffys}, {Tachibana}, {Sawada}, {Okazaki}, {Takano}, {Sakamoto},
  {Miura}, {Yano}, {Ireland}, {Yamada}, {Fujimoto}, {Kitazato}, {Namiki},
  {Arakawa}, {Hirata}, {Yurimoto}, {Nakamura}, {Noguchi}, {Yabuta}, {Naraoka},
  {Ito}, {Nakamura}, {Uesugi}, {Kobayashi}, {Michikami}, {Kikuchi}, {Hirata},
  {Ishihara}, {Matsumoto}, {Noda}, {Noguchi}, {Shimaki}, {Shirai}, {Ogawa},
  {Wada}, {Senshu}, {Yamamoto}, {Morota}, {Honda}, {Honda}, {Yokota},
  {Matsuoka}, {Sakatani}, {Tatsumi}, {Miura}, {Yamada}, {Fujii}, {Hirose},
  {Hosoda}, {Ikeda}, {Iwata}, {Kikuchi}, {Mimasu}, {Mori}, {Ogawa}, {Ono},
  {Shimada}, {Soldini}, {Takahashi}, {Takei}, {Takeuchi}, {Tsukizaki},
  {Yoshikawa}, {Terui}, {Nakazawa}, {Tanaka}, {Saiki}, {Yoshikawa}, {Watanabe},
  \& {Tsuda}}]{2022NatAs...6..214Y}
{Yada}, T., {Abe}, M., {Okada}, T., {et~al.} 2021, Nature Astronomy, 6, 214

\bibitem[{{Yamaguchi} {et~al.}(2023){Yamaguchi}, {Tomioka}, {Ito}, {Shirai},
  {Kimura}, {Greenwood}, {Liu}, {McCain}, {Matsuda}, {Uesugi}, {Imae},
  {Ohigashi}, {Uesugi}, {Nakato}, {Yogata}, {Yuzawa}, {Kodama}, {Hirahara},
  {Sakurai}, {Okada}, {Karouji}, {Nakazawa}, {Okada}, {Saiki}, {Tanaka},
  {Terui}, {Yoshikawa}, {Miyazaki}, {Nishimura}, {Yada}, {Abe}, {Usui},
  {Watanabe}, \& {Tsuda}}]{2023NatAs...7..398Y}
{Yamaguchi}, A., {Tomioka}, N., {Ito}, M., {et~al.} 2023, Nature Astronomy, 7,
  398

\end{thebibliography}

\begin{appendix}

\section{Proper elements of the primitive families}

\begin{figure}[h]
    \includegraphics[width=\columnwidth]{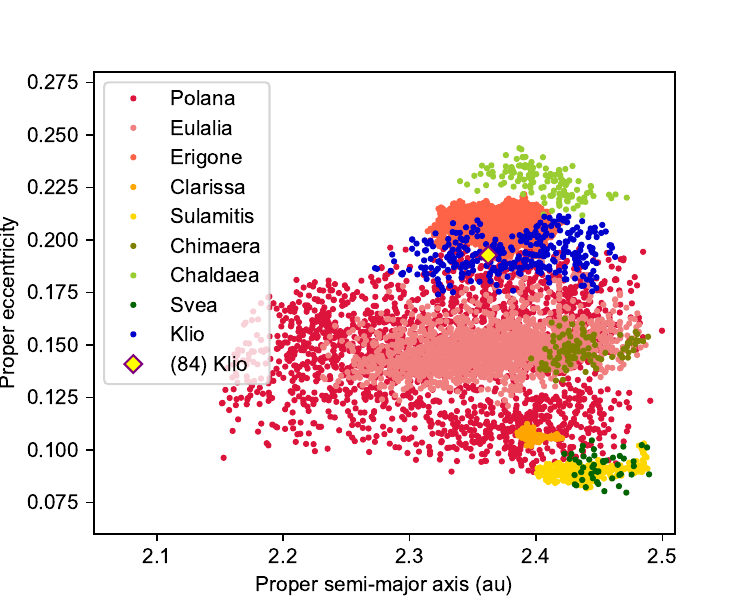}\\
    \caption{\textcolor{correc}{Proper eccentricity as a function of the proper semi-major axis of the low-albedo families in the inner main belt. For the Polana and Eulalia families, we used the definition from \citet{2013Icar..225..283W}, which was also used in \citet{2016Icar..266...57D}. For the other families, we applied the definition from \citet{2015aste.book..297N}. The yellow diamond marks the location of the asteroid (84) Klio.} }
    \label{fig:eccentricity_vs_semimajoraxis}
\end{figure}

\section{Klio's spectrum in the 3.4 and 3.8 $\mu$m spectral regions}

\begin{figure}[h]
    \includegraphics[width=\columnwidth]{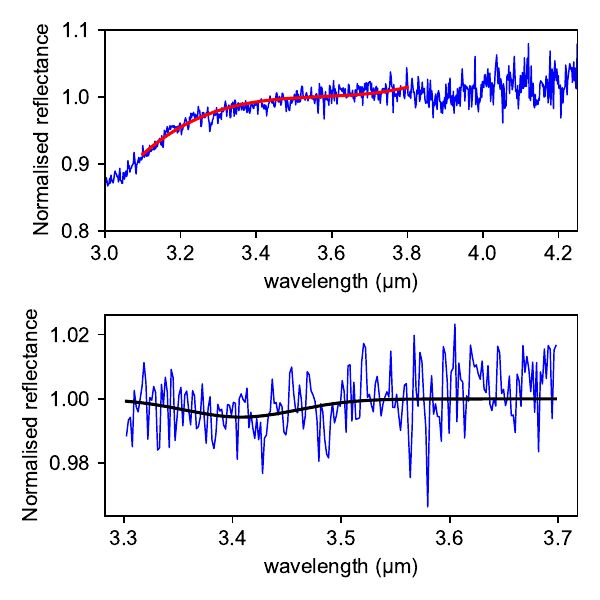}\\
    \caption{Top: polynomial continuum applied to the spectrum to analyse the 3.4 $\mu$m region. Bottom: Gaussian fit (in black) applied to the continuum-divided reflectance spectrum (in blue).}
    \label{fig:fit_orga}
\end{figure}

\begin{figure}[h]
    \includegraphics[width=\columnwidth]{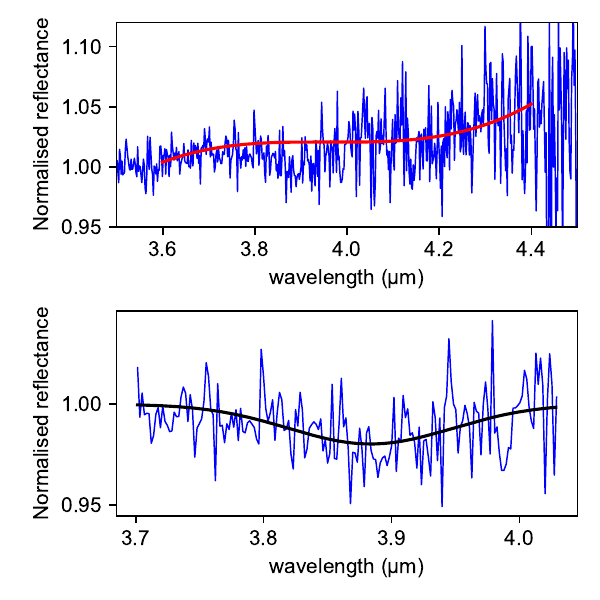}\\
    \caption{Top: polynomial continuum applied to the spectrum to analyse the 3.9 $\mu$m region. Bottom: Gaussian fit (in black) applied to the the continuum-divided reflectance spectrum (in blue).}
    \label{fig:fit_carbo}
\end{figure}

\section{Spectral comparison between Klio and CCs in the 0.7 $\mu$m region} \label{Sec:annex_spectral_comparison_VIS}

\begin{table}[!ht]
    \centering
    \caption{\textcolor{correc}{References of the RELAB spectra and samples used in this paper to compare Klio and the CCs in the 0.7 $\mu$m region.}}
    \begin{tabular}{ccc}
    \hline\hline\\[-3mm]
        name & RELAB Sample ID & RELAB Spectrum ID \\ 
        \hline\\[-3mm]
        Alais & MT-KTH-264 & C1MT264 \\
        Ivuna & MB-TXH-060 & C1MB60 \\ 
        Tagish Lake & MT-MEZ-01 & C1MT11 \\ 
        Bells & MB-TXH-053 & C1MB53 \\ 
        LAP 02277 & TB-MDD-355 & C1TB355 \\ 
        LAP 03786 & MP-TXH-219 & CAMP219 \\ 
        Murchison & MB-TXH-064-B & C3MB64 \\ 
        Murray & MB-TXH-056 & C1MB56 \\ 
        QUE 97990 & MT-JMS-219 & C1MT219 \\ 
        QUE 99038 & MP-TXH-233 & CAMP233 \\ 
        MIL 07700 & MP-TXH-229 & CAMP229 \\ 
        Al Rais & MT-KTH-265 & C1MT265 \\ 
        MIL 15328 & MP-TXH-258 & CAMP258 \\ 
        Mokoia & MT-KTH-272 & C1MT272 \\ 
        Vigarano & MB-TXH-059 & C1MB59 \\ 
        Felix & MT-KTH-298 & C1MT298 \\ 
        Isheyevo & TB-MDD-362 & C1TB362 \\ \hline
    \end{tabular}
    \label{tab:RELAB}

\end{table}

\begin{figure}
    \includegraphics[width=\hsize]{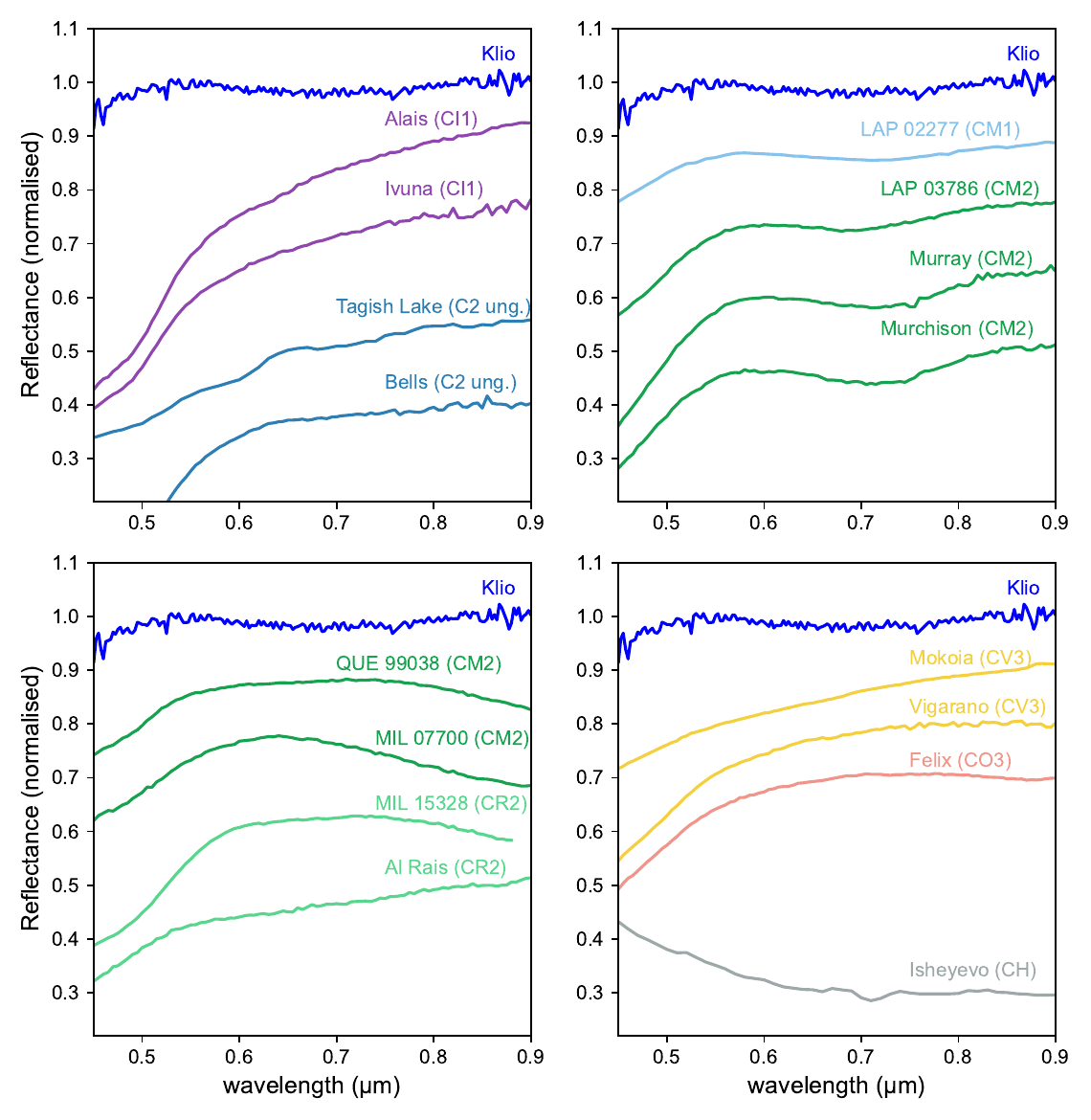}\\
    \caption{\textcolor{correc}{Spectral comparison of the 0.7 $\mu$m band between Klio \cite[SMASS-II spectrum,][]{2002Icar..158..106B} and several CCs from the RELAB database. All spectra have been normalised around 0.85 $\mu$m and have been shifted for better comparison.}}
    \label{fig:Klio_vs_alCCs_VIS}
\end{figure}

\newpage

\section{Aqueous alteration state of type 2 carbonaceous chondrites}
\label{sec:aq_alt}

\begin{table*}[!ht]
    \centering
    \begin{threeparttable}
    \caption{\textcolor{correc}{Supplementary information about the aqueous alteration state of CM carbonaceous chondrites used in this study.}}
    \begin{tabular}{cccccc}
    \hline\hline\\[-3mm]
    Name & Class\tnote{a} & Petrologic subtype\tnote{c} & Alexander Scale\tnote{d} & Howard Scale\tnote{f} \\[1mm]
    \hline\\[-3mm]
    Banten & CM2 & - & 1.7\tnote{d} & 1.6\tnote{g} \\ 
    Cold Bokkeveld & CM2 & 2.2\tnote{b,c} & 1.3\tnote{d} & 1.4\tnote{f} \\ 
    Crescent & CM2  & - & - & - \\
    LAP 02277 & CM1 & 2.1\tnote{b} / 2.0\tnote{c} & 1.4\tnote{g} & 1.2\tnote{g} \\
    LAP 03786 & CM2 & 2.2\tnote{b,c} & - & - \\ 
    MAC 02606 & CM2 & 2.1\tnote{b} & - & - \\ 
    MET 00639 & CM2 & 2.2\tnote{b} / 2.0-2.1\tnote{e} & 1.6\tnote{e} & 1.3\tnote{e} \\ 
    MIL 07700 & CM2 & 2.3\tnote{b} & - & - \\ 
    Murchison & CM2 & 2.5\tnote{c} & 1.6\tnote{d} & 1.5\tnote{f} \\
    Murray & CM2 & 2.4/2.5\tnote{c} & 1.5\tnote{d} & 1.5\tnote{f} \\ 
    QUE 97990 & CM2 & 2.6\tnote{b,c} & 1.7\tnote{d} & 1.6\tnote{f} \\ 
    QUE 99038 & CM2 & 2.4\tnote{b} & - & - \\ 
    \hline\\[-3mm]
    \end{tabular}
    \begin{tablenotes}
    \item[a] Class used by \citet{2013M&PS...48.1618T,2019Icar..333..243T}
    \item[b] values in \citet{2013M&PS...48.1618T}
    \item[c] Definition of the scale / values in \citet{2007GeCoA..71.2361R}
    \item[d] Definition of the scale / values in  \citet{2013GeCoA.123..244A}
    \item[e] values found in \citet{2019LPICo2189.2115D}
    \item[f] Definition of the scale / values in \citet{2015GeCoA.149..206H} 
    \item[g] values found in \citet{2021GeCoA.310..240K}
    \end{tablenotes}
    
    \end{threeparttable}
    \label{tab:aq_alt}
\end{table*}

\end{appendix}

\end{document}